\documentclass[11pt]{article}
\pdfoutput=1
\usepackage[totalwidth=480pt, totalheight=680pt]{geometry}

\usepackage[latin1]{inputenc}

\usepackage{hyperref}

\usepackage{amsmath}
\usepackage{amssymb}
\usepackage{latexsym}
\usepackage{bbm}

\usepackage[above,below]{placeins}
\usepackage[title,titletoc]{appendix}

\usepackage{ifpdf}

\ifpdf
\usepackage[pdftex]{graphicx}
\else
\usepackage{graphicx}
\fi

\graphicspath{{curves/}}

\title{Efficient Pricing of CPPI using Markov Operators}
\author{ Louis Paulot and Xavier Lacroze \\ \emph{Sophis} }

\date{November 28, 2008}

\begin{document}

\ifpdf
\DeclareGraphicsExtensions{.pdf, .jpg, .tif}
\else
\DeclareGraphicsExtensions{.eps, .jpg}
\fi

\thispagestyle{empty}


\vspace*{3cm}

\begin{center}
\begin{Large}
\textbf{Efficient Pricing of CPPI using Markov Operators}
\end{Large}

\vspace{10mm} {\bf Louis Paulot and Xavier Lacroze}

\vspace{7mm}
\emph{Sophis Technology
\\
24--26 place de la Madeleine, 75008 Paris, France}

\vspace{5mm}
{\ttfamily louis.paulot@sophis.net\\xavier.lacroze@sophis.net}

\end{center}

\vspace{4cm}
\hrule
\begin{abstract}

Constant Proportion Portfolio Insurance (CPPI) is a strategy designed to give participation in a risky asset while protecting the invested capital. Some gap risk due to extreme events is often kept by the issuer of the product: a put option on the CPPI strategy is included in the product. In this paper we present a new method for the pricing of CPPIs and options on CPPIs, which is much faster and more accurate than the usual Monte-Carlo method. Provided the underlying follows a homogeneous process, the path-dependent CPPI strategy is reformulated into a Markov process in one variable, which allows to use efficient linear algebra techniques. Tail events, which are crucial in the pricing are handled smoothly. We incorporate in this framework linear thresholds, profit lock-in, performance coupons... The American exercise of open-ended CPPIs is handled naturally through backward propagation. Finally we use our pricing scheme to study the influence of various features on the gap risk of CPPI strategies.

\end{abstract}
\hrule

\vspace{\stretch{1}}

\pagebreak

\setcounter{tocdepth}{2}
\tableofcontents

\pagebreak

\section{Introduction}

Constant Proportion Portfolio Insurance is a dynamic strategy designed to give participation in some risky asset while protecting the invested capital \cite{merton1970oca}. This is achieved by rebalancing periodically between a risk-free asset (Zero-Coupon bond) and a risky asset(share, index, fund, fund of funds\ldots). In the simplest form, if the underlying asset has no jumps and if one can rebalance continuously, the final payoff depends in a deterministic way on the risky underlying. However both hypothesis are strong and do not fit real market conditions. Relaxing them, the strategy is not as efficient: there is a small chance of not recovering the whole capital invested at maturity. This gap risk may be hold by the issuer, so that the principal is really guaranteed to the investor. In this case, there is an option included in the product, which must be priced and hedged. With a discrete rebalancing scheme, there is a closed formula for this option price if the underlying follows a Black-Scholes diffusion \cite{balder:ecs}. Moreover, this formula can be generalized to jump-diffusion models, and more generally to Levy processes. However, this works only for an idealized CPPI product where there are no spreads on the risk-free and financing rates, no fees, a natural bond floor... Unfortunately, real CPPIs always have such features, which prevent from using the closed formula. As CPPIs are very path dependent, they are usually priced through Monte-Carlo simulations (see \cite{boulier} for an example). Extreme value theory has also been used to estimate the gap risk of such products \cite{bertrand2002pie}.

Monte-Carlo pricing is perfectly suited to path dependence but the dimension of the problem is generally quite large. For example, a monthly CPPI defined on a single risky asset with monthly rebalancing and a maturity of 20 years requires 240 fixings. This means a 240-dimensional Monte-Carlo integration. Furthermore, the tails of the distributions play important roles in the pricing. First, the lower tail gives the gap risk part. It must be computed with enough precision to produce a reasonable Put price and an accurate Delta. The upper tail is important to reach the correct mean value of the CPPI strategy. The payoff distribution is close to a shifted lognormal distribution with high leverage: most trajectories will end far below the mean and will be compensated by few very high terminal values. Control variate can solve part of the convergence problems for the strategy mean value, by comparing the payoff of the real CPPI to the payoff of a self-financing CPPI. However that does not solve all convergence issues: in particular the optional part, the gap risk, is not smoothed by basic control variate. The high path dependence and barrier-like structure of the strategy makes the Put price and its Delta converge very slowly. Moreover, more and more CPPIs are open-ended: the investor can exit at any time. The Monte-Carlo method is not well suited to this American feature.

We propose here an alternative way for pricing CPPI strategies and derivatives. Instead of simulating trajectories, transition probabilities are computed. This takes care smoothly of distribution tails and produces smooth greeks. Initially, there are four correlated variables to follow: the risky asset, the CPPI strategy value, the risky asset exposure and the guaranteed amount. In many situations, the problem can be reformulated in such a way that it becomes one-dimensional and Markovian. Instead of diffusing the underlying, the CPPI index value is evolved. Transition matrices between two rebalancing dates are computed. From the Markov property, the final distribution can be easily computed from the starting point and one can even compute the full transition matrix. The main advantages of this technique are its speed and the smoothness of greeks. For example in the simplest cases, only a few milliseconds are needed to get an accurate price. More complex cases with profit lock-in, artificial cushion or coupons take more time but never more than a few seconds with correct approximation. As the pricing is done through backward propagation, products with American optionnality such as open-ended CPPIs are priced naturally.

After setting the model assumptions in section \ref{section-hypothesis}, our pricing method is described in section \ref{section-pricing}. Features which deserve a special treatment are considered in \ref{section-pricing-features}: performance coupons, profit lock-in, conditional rebalancing\ldots We discuss the influence of the discretization scheme and how to keep under control the numerical artifacts arising from the discretization process in section \ref{section-numerical}. Numerical results are presented in the two last sections. Convergence results are shown and compared to a Monte-Carlo pricing in section \ref{section-convergence}. Then we illustrate the efficiency of the pricing by studying the impact of various CPPI features on the final distribution and especially on the gap risk in section \ref{section-features-model}. Finally, two appendices give useful formulas for the jump-diffusion model we use and for vanilla CPPI prices.

\section{Assumptions}
\label{section-hypothesis}

\subsection{Numéraire choice}

We consider two cases.
\begin{enumerate}
\item \emph{Deterministic interest rates.} The numéraire is the monetary unit. All features described in this paper can be handled in this case.

\item \emph{Stochastic interest rates.} In this case we choose the forward risk-neutral measure: the numéraire is a Zero-Coupon bond of maturity larger than the CPPI maturity. The assets which are diffused are of the form $\frac{S_t}{ZC(t,T)}$. The stochastic nature of interest rates can be forgotten provided every quantity can be expressed simply in term of this numéraire. Linear threshold and fixed rate threshold are therefore excluded, because they have "absolute" value in monetary unit and are not multiple of a Zero-Coupon bond. With this numéraire, the threshold would become stochastic. It is not possible to reduce the diffusion to a single asset, one has really to model two assets. To be able to use stochastic interest rates in this framework, the threshold must be defined as a Zero-Coupon curve, with the possibility of adding some spread over the risk-free rate.

\end{enumerate}

Formally, we will write deterministic interest rates, which may come from stochastic rates in forward risk-neutral measure. Just remind that this latter case is not compatible with linear or fixed rate threshold.

\subsection{Underlying process}

In order to be able to model directly the CPPI strategy value without following the underlying asset in addition, we make the assumption that the risky underlying evolution is homogeneous: $\Delta S / S$ does not depend on $S$. However it can depend on time-dependent parameters.

Black-Scholes diffusion will work. Similarly, models with jumps which respect the Levy property (independent increments) fit naturally in the framework. In the appendix, we describe the Kou model with exponentially distributed positive and negative jumps but other jump distributions can be used (Merton's for instance). The choice of the model depends on the context: for pricing and hedging a market calibrated model has to be used, whereas for portfolio optimization or risk management historical probability can be used \cite{cont:cpp}.

The hypothesis on the underlying evolution mainly excludes local volatility models. Stochastic volatility models can be incorporated in this framework but the volatility must be included in the variables space, so a few volatility regimes would be more appropriate numerically than a continuous volatility diffusion \cite{albanese2006oma}. If the CPPI value is discretized on a grid with $N$ points, adding $N_\sigma$ points for volatility will replace $N \times N$ transition matrices with $N N_\sigma \times N N_\sigma$ matrices. Starting from $N \sim 200$, a low $N_\sigma$, between 3 and 5, will keep the matrices tractable: $N N_\sigma \sim 1000$.

\section{CPPI Pricing}
\label{section-pricing}

In a direct approach, four variables have to be modeled: the risky asset value, the CPPI index value, the risky asset weighting and the guaranteed amount. In most cases the problem can be expressed as a Markov chain on one single process, the CPPI index value rescaled by the threshold value. Transition operators between two rebalancing dates can be constructed. Their multiplication gives the transition operator over the whole CPPI life which is used to price the CPPI strategy or options on the strategy.

In this section we deal with the basic case where there is no profit lock-in, coupons or American exercise. These features are addressed in the following section.

\subsection{Definitions}

\subsubsection{Self-financing CPPI}

The CPPI strategy starts at $t_0 = 0$ by investing a nominal amount that we take equal to 1 to simplify formulas and ends at $t_n = T > 0$. The portfolio is rebalanced at times $t_i$ for $i = 0,\dots,n-1$ such that $t_0  < t_1 < \ldots < t_n$. The value of the portfolio at time $t_i$ is denoted $C_i = C(t_i)$. Similarly, we write $S_i$ for the price of the risky asset on that date and $F_i$ for its forward price with maturity $T$. $B_i$ denotes the value of the risk-free asset: for a self-financing CPPI, $B_i$ is given by the money market:
\begin{equation}
B_i \propto \text{ZC} (t_i, T)
\end{equation}
where $\text{ZC}(\cdot,T)$ stands for the Zero-Coupon of maturity $T$.

In a first step, the guarantee\footnote{Note that whereas the previously defined processes were in fact continuous, the ones we now introduce are only relevant at times $t_i$.} $G_i$ is constant and such that $G_i = G = 1$ for all $i$. The threshold $H_i$ is a deterministic function of the guaranteed amount $G_T$ and of the time $t_i$. The natural choice is
\begin{equation*}
H_i = G \ \text{ZC} (t_i, T)
\end{equation*}

The rebalancing criterion at $t_i$ states that a stochastic proportion $W_i = w(C_i,H_i)$ of the portfolio value should be invested in the risky asset $S$ while a proportion $1-W_i$ is invested in the risk-free asset $B$. The proportion $W_i$ is called the risky asset weighting and is usually chosen of the form
\begin{equation}
W_i = w(C_i,H_i) = m \left( \frac{C_i - H_i}{C_i} \right)^+ = m \left( \frac{C_i/H_i - 1}{C_i/H_i} \right)^+
\label{eq-standard-exposure}
\end{equation}
where $m > 0$ is the multiplier of the CPPI. We could also include minimum and/or maximum exposures, a cushion limit which triggers full investment in the riskfree asset or any other feature, even time dependent, the only restriction being that $W_i$ should be a non negative deterministic function $W_i = w(X_i)$ of the ratio
\begin{equation*}
X_i = \frac{C_i}{H_i}
\end{equation*}
More advanced strategies could be handled at the cost of also following the variable $W_i$ which makes our problem 2 dimensional (see section \ref{section-conditionalrebalancing}).

From now on, we refer to the CPPI we have just described as a self-financing CPPI. Closed analytical formulas for this special case are derived in appendix \ref{appendix-closed-formula}.

\subsubsection{Spreads, artificial thresholds, fees\ldots}

We introduce the following possibly time-dependent spreads, expressed in continuous mode.\footnote{If rates and spreads are defined in some other compounding mode, as linear mode for example, the continuous spread will slightly depend on the risk-free interest rate $r$.}
\begin{itemize}
\item A spread $s_+$ on the risk-free asset, it is usually non-positive.
\item A spread $s_-$ on the financing component (i.e.\ when the risky asset weighting is greater than 1). Usually non-negative.
\end{itemize}
This means that over the period from $t_i$ to $t_{i+1}$, we have
\begin{equation*}
\frac{B_{i+1}}{B_i} = e^{\int_{t_i}^{t_{i+1}} \left( r + s_\pm \right) \, \mathrm{d}t}
\end{equation*}
where $s_\pm = s_+ \mathbbm{1}_{W_i \leq 1} + s_- \mathbbm{1}_{W_i > 1}$.

We also introduce a (possibly time-dependent) spread $s_H$ for the threshold rate, such that
\begin{equation*}
\frac{H_{i+1}}{H_i} = e^{\int_{t_i}^{t_{i+1}} \left( r + s_H \right) \, \mathrm{d}t}
\end{equation*}
If $s_H$ is not zero, the variable $X_i = C_i/H_i$ is no longer the forward value of the CPPI. As the threshold is only defined at rebalancing dates, a time dependent spread can be used to "simulate" a linear threshold. As for the risky asset weighting, the sole restriction on $H_i$ is that is a positive time dependent deterministic function of the guarantee such that the threshold at maturity $H_T$ equals to the guarantee.

Finally, we introduce some managing fees depending on the index value at the beginning of the period\footnote{One could do the same with fees depending on the value at the end of the period.}, the risky asset weighting\dots We consider here several types of fees (possibly time dependent):
\begin{itemize}
\item proportional fees $f_p C_i$
\item proportional fees replaced with some discounted defeasance rate $f_d$ when all the investment is invested in the risk-free asset:
    $\left( f_p \mathbbm{1}_{W_i > 0} + f_d \mathbbm{1}_{W_i = 0} \right) C_i$
\item fees on the risky investment $f_r W_i C_i$
\item fixed fees $f_f$
\end{itemize}
The sum of all fees is written as $f(C_i, W_i) C_i$ for convenience.

\subsection{Transition matrix over one period}

The pricing is performed by computing the transition matrix of the CPPI value between the computation date to its maturity. For this operator to be computed, the first step is the computation of transition matrices over every period.

\subsubsection{Self-financing CPPI}

We consider here the simplest case where there are neither spreads nor fees ($s_\pm = s_H = f = 0$) and no profit lock-in ($G_i=G=1$).
Consider the "rescaled" CPPI value $X_i = C_i/H_i$. It is a Markov process in discrete time and we construct a local (e.g.\ from one rebalancing date $t_i$ to the following one) transition operator $M_i$ as follows.
The CPPI variation between $t_i$ and $t_{i+1}$ is given by
\begin{equation*}
C_{i+1} = w(X_i) \, C_i \, \frac{S_{i+1}}{S_i} + \Big( 1-w(X_i) \Big) C_i \, \frac{B_{i+1}}{B_i}
\end{equation*}
In terms of the variables $X_i$ and denoting by $F_i$ the value of the forward on the risky asset with maturity $T$, this looks even simpler:
\begin{equation}
\frac {X_{i+1}}{X_i} = 1 + w(X_i) \left( \frac{F_{i+1}}{F_i} - 1 \right)
\label{eq-SF-oneperiod}
\end{equation}

The evolution operator for the period $(t_i,t_{i+1})$ is defined as
$$
\mathcal M_i (x,y) = \mathbb{E} \left[ \delta \left( X_{i+1} - y \right) \, \big\vert \, X_i = x \right]
$$
i.e.\ it is the density of $X_{i+1}$ conditionally on $X_i = x$. We can compute $\mathcal M_i$ through the previous equation as
\begin{equation*}
\mathcal M_i (x, y) = \frac{\partial}{\partial y} \mathbb{P} \left[ X_{i+1} \leq y \, \big\vert \, X_i = x \right] = \frac{\partial}{\partial y} \mathbb{P} \left[ \frac{F_{i+1}}{F_i} \leq L(x,y) \right]
\end{equation*}
with
\begin{equation*}
	L(x,y) = 1 + \frac{ y - x }{w(x)\,x}
\end{equation*}

\subsubsection{General case}

The only difference with the previous case is that the computation are a bit nastier. The value of the CPPI portfolio at time $t_{i+1}$ after the fees were paid is
\begin{equation*}
\frac{C_{i+1}}{C_i} = w(X_i) \frac{S_{i+1}}{S_i} + \left( 1-w(X_i) \right) \frac{B_{i+1}}{B_i} - f(C_i,w(X_i))
\end{equation*}
In term of the variable $X$ and the forward $F$, this reads
\begin{equation*}
\frac{X_{i+1}}{X_i} = \frac{C_{i+1}}{C_i} \frac{H_i}{H_{i+1}} = e^{-\int_{t_i}^{t_{i+1}} s_H \, \mathrm{d}t} \left( w(X_i) \frac{F_{i+1}}{F_i} + \left( 1-w(X_i) \right) \, e^{\int_{t_i}^{t_{i+1}} s_\pm \, \mathrm{d}t} - e^{-\int_{t_i}^{t_{i+1}} r \, \mathrm{d}t} f(C_i,w(X_i)) \right)
\end{equation*}
We invert this relation to get the variation of $F$ conditionally on $X_i$ needed to achieve some value of $X_{i+1}/X_i$ as
\begin{equation*}
\frac{F_{i+1}}{F_i} = \frac{1}{w(X_i)} \left( e^{\int_{t_i}^{t_{i+1}} s_H \, \mathrm{d}t} \frac{X_{i+1}}{X_i} - (1-w(X_i)) \, e^{\int_{t_i}^{t_{i+1}} s_\pm \, \mathrm{d}t} + e^{-\int_{t_i}^{t_{i+1}} r \, \mathrm{d}t} f(C_i,w(X_i)) \right)
\end{equation*}

As above, we get the transition kernel between times $t_i$ and $t_{i+1}$ as
\begin{equation*}
\mathcal M_i (x, y) = \frac{\partial}{\partial y} \mathbb{P} \left[ \frac{F_{i+1}}{F_i} \leq L(y,x) \right]
\end{equation*}
where
\begin{equation*}
L (x,y) = \frac{1}{w(x)} \left( e^{\int_{t_i}^{t_{i+1}} s_H \, \mathrm{d}t} \frac{y}{x} - \big(1-w(x)\big) e^{\int_{t_i}^{t_{i+1}} s_\pm \, \mathrm{d}t} + e^{-\int_{t_i}^{t_{i+1}} r \, \mathrm{d}t} f(x,w(x)) \right)
\end{equation*}

In the numerical implementation, we make use of the expected value of $X_{i+1}$ conditionally on $X_i$ and $X_{i+1} \leq y$. We quote here the result
\begin{multline*}
\mathbb{E} \left[ X_{i+1} \mathbbm{1}_{X_{i+1} < y} \, \big\vert \, X_i = x \right] =
\\
x \, e^{ -\int_{t_i}^{t_{i+1}} s_H \, \mathrm{d}t} \left(
w(x) \, \mathrm P_2 + \left( \big( 1-w(x) \big) e^{\int_{t_i}^{t_{i+1}} s_\pm \, \mathrm{d}t} - e^{-\int_{t_i}^{t_{i+1}} r \, \mathrm{d}t} f(x,w(x)) \right) \mathrm P_1 \right)
\end{multline*}
which is expressed in term of the cumulative $\mathrm P_1 = \mathbb{P} \left[ \frac{F_{i+1}}{F_i} < L(x,y) \right]$ and the "conditional" expected value $\mathrm P_2 = \mathbb{E} \left[ \frac{F_{i+1}}{F_i} \mathbbm{1}_{\frac{F_{i+1}}{F_i} < L(x,y)} \right]$ of $\frac{F_{i+1}}{F_i}$ (see appendix \ref{section-model}).

\subsection{Transition matrix over the CPPI life}
\label{section-multiperiod}


The evolution kernel over the whole CPPI life is defined as
\begin{equation*}
\mathcal K (x,y) = \mathbb{E} \left[ \delta( X_n - y ) \, \big\vert \, X_0 = x \right]
\end{equation*}
allows to compute the expected value of a payoff $P(C_n)$ (as there is no profit lock-in, here we have $G_i=G=1$ and thus $X_n = C_n$) as
\begin{equation*}
\mathbb{E} \left[ P(X_n) \, \big\vert \, X_0 \right]  = \int \mathrm{d}y \, \mathcal K \left( X_0,y \right) P(y)
\end{equation*}
In operator terms, this is the result of the action of the operator $\mathcal K$ on the function $P$, the result being evaluated at $X_0$:
\begin{equation*}
\mathbb{E} \left[ P(X_n) \, \big\vert \, X_0  \right] = \left( \mathcal K \cdot P \right) (X_0)
\end{equation*}
Finally we define the partial evolution operators between times $t_i$ and $t_j$ as
\begin{equation*}
\mathcal K_{ij} (x,y) = \mathbb{E} \left[ \delta( X_j - y ) \, \big\vert \, X_i = x \right]
\end{equation*}

The CPPI process is Markovian: the evolution operator over all periods is given by the product of evolution operators over single periods:
\begin{equation*}
\mathcal K_{ij} = \mathcal M_i \mathcal M_{i+1} \ldots \mathcal M_{j-1}
\end{equation*}
where $\mathcal M_n$ is the transition operator over the period $(t_i,t_{i+1})$ we considered in the previous section.
The total evolution is given by the operator
\begin{equation}
\label{eq-whole-kernel}
\mathcal K = \mathcal K_{0n} = \mathcal M_0 \mathcal M_1 \ldots \mathcal M_{n-1}
\end{equation}
If all local evolution operators $\mathcal M_i$ are equal to a common operator $\mathcal M$, this reduces to the power
\begin{equation*}
\mathcal K = \mathcal K_{0n} = \mathcal M^n
\end{equation*}

The fair value today of the product is the expected value
\begin{equation*}
(\mathcal K P)(X_0) = \left(\mathcal M_0 \mathcal M_1 \ldots \mathcal M_{n-1} P \right)(X_0)
\end{equation*}

\subsection{Greeks}

Before the CPPI starts, and hence before the first rebalancing of the portfolio, there is of course no sensitivity to the risky asset price. Gamma and delta will be identically null until the strategy really starts. At a time $t$ such that $t_0 \leq t < t_1$, consider the spot $S_t$ and the forward $F_t$ of the underlying asset. We can decompose equation \eqref{eq-SF-oneperiod} as
\begin{equation*}
\frac {X_1}{X_0} = w(X_0) \left( \frac{F_1}{F_t} \frac{F_t}{F_0} - 1 \right) + 1
\end{equation*}
and one gets that transition matrix for the remaining time of the first period is given by $\mathcal M_0 (x,y) = \frac{\partial}{\partial y} \mathbb{P} \left[ \frac{F_1}{F_t} \leq L_0(x,y) \right]$
with
\begin{equation*}
	L_0(x,y) = \frac{F_0}{F_t} \left( \frac{ y - x }{w(x) \, x}+ 1 \right)
\end{equation*}
One can then compute the derivative of the kernel $\mathcal M_0 (x,y)$ with respect to $F_t$ (or the spot $S_t$) and get the sensitivity of the price to the risky asset as
\begin{equation*}
\frac{\partial}{\partial F_t} \mathbb{E} \left[ P(X_n) \, \big\vert \, X_0 \right] = \int \mathrm{d}y \, \frac{\partial \mathcal K \left( X_0,y \right)}{\partial F_t} P(y)
\end{equation*}
with, using \eqref{eq-whole-kernel},
\begin{equation*}
\frac{\partial \mathcal K \left( X_0,y \right)}{\partial F_t}  = \frac{\partial \mathcal M_0}{\partial F_t} \mathcal M_1 \ldots \mathcal M_{n-1}
\end{equation*}
as only the first kernel depends on the spot of the risky underlying.

\subsection{Gap Risk}

As sold CPPI are usually guaranteed, we define three useful measures of risk in that context: the put price (with a strike equal to the guaranteed amount at maturity), the gap proportion and the expected gap loss. The gap proportion is defined as the probability that the final value of the CPPI will be below the threshold and the expected gap loss as the expected loss knowing that the CPPI ends up below the threshold. The gap proportion is the price of the undiscounted digital put and expected gap loss is the quotient of the undiscounted put price by the gap proportion.

\section{Additional features}
\label{section-pricing-features}

CPPI products have often additional features: coupons, profit lock-in, rebalancing conditions... We show in this section how they can be handled in the pricing process.

\subsection{Coupons}

\subsubsection{Definitions}

The investor might want to be remunerated for the performance but not wait till the CPPI maturity. At times $t_{c(I)}$, the investor will then receive money in the form of coupons. To simplify notation, we use capital indices to designate coupon dates: $X_I$ means $X_{c(I)}$ where $c(I) \in \{0, \ldots, n\}$ and $I \in \{0,\ldots,n_c\}$ with $n_c$ the number of coupons.
Two types of coupons have been devised:
\begin{enumerate}
	\item \emph{Fixed coupon}. Fixed coupons are mostly (time dependent) fixed fees except for the fact that they are paid to the investor and hence negative.

	\item \emph{Performance coupon}. Periodically some proportion $q$ of the performance over the last period is paid to the investor. If $C^-_I$ is the value of the CPPI at time $t_I$ immediately before the coupon is detached, the value $K_I$ of the coupon paid at for the period $(I-1,I)$ is computed as
\begin{equation*}
K_I = q \left( C^-_I - C_{I-1} \right)^ + \mathbbm{1}_{W_{I-1}>0}
\end{equation*}
and
\begin{equation*}
C_I = C^-_I - K_I
\end{equation*}
Setting $X^-_I = C^-_I / H_I$, the previous equation reads
\begin{equation*}
X_I = X_I^- - q \left( X^-_I - \frac{H_{I-1}}{H_I} X_{I-1} \right)^+ \mathbbm{1}_{W_{I-1}>0}
\end{equation*}
If no coupon is paid $X_I = X^-_I$. In the case a coupon is paid, $K_I > 0$ and
\begin{equation*}
X^-_I = \frac{1}{1-q} \left( X_I - q \frac{H_{I-1}}{H_I} X_{I-1} \right)
\end{equation*}

It is also useful to compute the value of the coupon in proportion of the guarantee as a function of $X^-_I$ and $X_I$:
\begin{equation*}
\frac{K_I}{G} = q \left( X^-_I e^{- \int_{t_I}^T r + s_H \, \mathrm{d}t} - X_{I-1} e^{- \int_{t_{I-1}}^T r + s_H \, \mathrm{d}t} \right)^ + \mathbbm{1}_{W_{I-1}>0}
\end{equation*}

\end{enumerate}

\subsubsection{Pricing with performance coupons}

Fixed coupons do not exhibit particular difficulties as they are essentially (negative) fixed fees, hence we concentrate here on the pricing of performance coupons.

Consider $\mathcal M_{I-1} = \mathcal K_{I-1,I}$ be the transition operator, without any coupon, between $t_{c(I-1)}$ and $t_{c(I)}$. The expected value of the coupon capitalized up to maturity $T$, conditionally on $X_{I-1}$, is given by
\begin{align*}
	c_I (x) & = \frac{1}{\text{ZC}(t_I,T)} \mathbb{E} \left[ \frac{K_I}{G} \, \big\vert \, X_{I-1} = x \right] \\
	& = \frac{q \mathbbm{1}_{w(x)>0}}{\text{ZC}(t_I,T)} \int \mathrm{d}z \, \mathcal M_{I-1} (x, z) \left( z e^{- \int_{t_I}^T r + s_H \, \mathrm{d}t} - x e^{- \int_{t_{I-1}}^T r + s_H \, \mathrm{d}t} \right)^+
\end{align*}
In order to account for the payment of the coupon, we define a rescaled operator $\widetilde {\mathcal M}$:
\begin{equation*}
\widetilde{\mathcal M}_{I-1} (x, y) = \int \mathrm{d}z \, \delta \left( z - y - q \mathbbm{1}_{w(x)>0} \left( z - x e^{\int_{t_I}^{t_{I+1}} r + s_H \, \mathrm{d}t} \right)^+ \right) \mathcal M_{I-1} (x, z)
\end{equation*}
By construction this operator satisfies
\begin{equation*}
\widetilde {\mathcal M}_{I-1} (x,y) = \mathbb{E} \left[ \delta( X_I - y ) \, \big\vert \, X_{I-1} = x \right]
\end{equation*}
which enables us to write $\mathcal K = \mathcal K_{0N_c}$ for the pricing kernel, with
\begin{equation*}
\mathcal K_{IJ} = \widetilde {\mathcal M}_I \ \widetilde {\mathcal M}_{I+1} \ \ldots \ \widetilde{\mathcal M}_{J-1}
\end{equation*}
However, to get the correct price for the strategy, the value of the coupons to be paid must be included. The total value of all coupons is
\begin{equation*}
\sum_{I=1}^{n_c} \frac{1}{\text{ZC}(t_I,T)} \mathbb{E} \left[ K_I \, \big\vert \, X_{0} \right] = G \sum_{I=1}^{n_c} \frac{1}{\text{ZC}(t_I,T)} \mathbb{E} \left[ \frac{K_I}{G} \, \big\vert \, X_{0} \right] = G \sum_{I=1}^{n_c} (\mathcal K_{0,I-1} \cdot c_I) (X_0)
\end{equation*}

In fact, this hides a semi-direct product structure which, numerically speaking, will be handled as matrix multiplication in section \ref{section-numerical}. For a pair constituted of an operator $\mathcal O$ and a function $f$, the action of the pair on a function $g$ is given by
\begin{equation*}
\left( \left( \mathcal O, f \right) \cdot g \right) (x) = \left( \mathcal O \cdot g \right) (x) + f(x)
\end{equation*}
The (semi-direct) product of two pairs is
\begin{equation*}
\left( \mathcal O_1, f_1 \right) \ \left( \mathcal O_2, f_2 \right) = \left( \mathcal O_1 \mathcal O_2, \ \mathcal O_1 \cdot f_2 + f_1 \right)
\end{equation*}
We claim that the price of strategy with payoff of the form $G_T P(X_T)$ including the coupons is given by
\begin{equation*}
G \left( \widetilde{\mathcal M}_{0}, c_0 \right) \left( \widetilde{\mathcal M}_{1}, c_1 \right) \ldots \left( \widetilde{\mathcal M}_{n_c}, c_{n_c} \right) \cdot P
\end{equation*}
taken at $X_0$ with the convention $c_0 (x) = 0$.

\subsection{Profit lock-in}
\label{section-LI}

\subsubsection{Definitions}

In order to lock some part of the profit already made, the guarantee $G$ is turned into a stochastic process $G_i$.

We denote by a minus exponent ($X^-_I$) the value of a variable before applying the lock-in.
We parametrize the lock-in by a multiplicative factor $f$ which relates $X_{I+1}$ to $X^-_{I+1}$ through
\begin{equation*}
\frac{X^-_{I+1}}{X_{I+1}} = \frac{G_{I+1}}{G_{I}} = f_{I+1}(X_{I}, X^-_{I+1})
\end{equation*}

There are two main kinds of profit lock-in:
\begin{enumerate}
\item \emph{Periodic lock-in}. Periodically (every year for example) some fixed proportion $p$ of the performance since the last lock-in is added to the guarantee. As in the case of coupons, we use capital indices to designate lock-in dates: $X_I$ means $X_{l(I)}$ where $l(I) \in \{0, \ldots, n\}$ and $I \in \{0,\ldots,n_l\}$ with $n_l$ the number of lock-ins. The new guarantee value is given by
\begin{equation*}
G_{I+1} = G_{I} + p \left( C_{I+1} - C_{I} \right)^+
\end{equation*}
Between two lock in, the guaranteed amount is constant.

We denote by $H^-_{I+1} = H_{I+1} \ G_{I+1} / G_I$ the value of threshold immediately before the profit lock in. Similarly, we define $X^-_{I+1} = C_{I+1} / H^-_{I+1}$.
In term of $X_I$ and $X^-_{I+1}$ the guaranteed amount is rescaled by a factor
\begin{equation*}
f_{I+1}(X_I,X^-_{I+1}) =  \frac{G_{I+1}}{G_{I}} =  1 + p \left( X^-_{I+1} e^{- \int_{T_{l(I+1)}}^T \left( r + s_H \right) \, \mathrm{d}t} - X_I e^{-\int_{T_{l(I)}}^T \left( r + s_H \right) \, \mathrm{d}t} \right)^+
\end{equation*}

\item \emph{Continuous lock-in}. On rebalancing dates, some proportion $p$ of the highest CPPI value attained is guaranteed. The new guarantee is (in fact, in this case, $l(I) = I$ since lock-ins happen at every rebalancing dates)
\begin{equation*}
G_{I+1} = \max \left( G_{I}, \ p C_{I+1} \right)
\end{equation*}
The multiplicative factor in this case is a function of $X^-_{I+1}$ alone and is given by
\begin{equation*}
f_{I+1}(X^-_{I+1}) = \max \left( 1, \ p X^-_{I+1} e^{- \int_{t_{I+1}}^T \left( r + s_H \right) \, \mathrm{d}t} \right)
\end{equation*}
\end{enumerate}

\begin{figure*}[ht]
\centering
\includegraphics[width=0.75\textwidth]{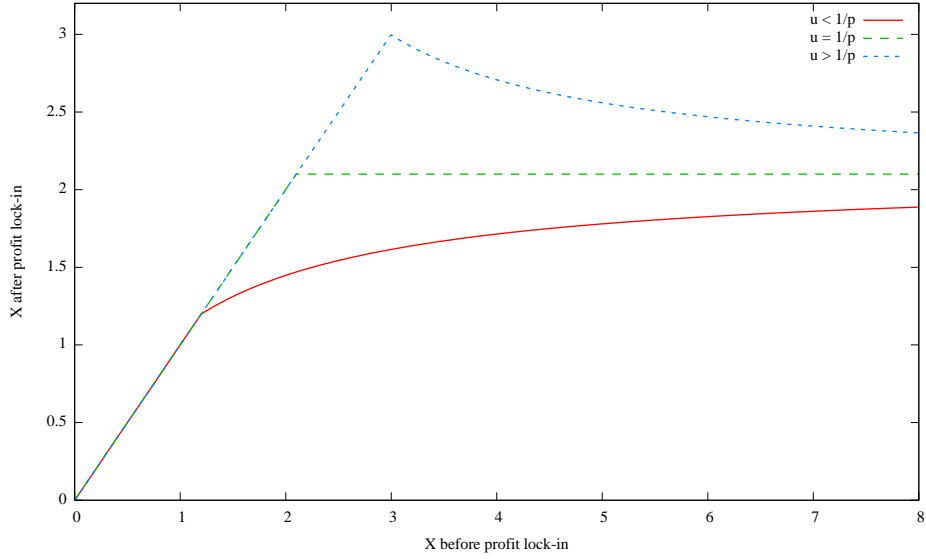}
\caption{\emph{$X_{I+1} = X^-_{I+1} / f_I (X_I,X^-_{I+1})$ as a function of $X^-_{I+1}$ in the case of a periodic profit lock-in with $p=0.5$ for three values of $u = X_I e^{-\int_{I}^{T} r + s_H \, \mathrm{d}t}$: $u=1.14$, $u=2$ and $u=2.85$. The threshold value at $I$ is equal to $0.95$.}}
\label{factorf}
\end{figure*}

To get some intuition on what happens between $X^-_{I+1}$ and $X_{I+1}$ in the case $X_I > 1$, see Figure~\ref{factorf}. The scale is in unit of $e^{- \int_{t_{I+1}}^T r + s_H \, \mathrm{d}t}$. For periodic profit lock-in, there are three different regimes based on whether one started close to threshold or far above, the last regime being merely a transition between the first two. In the case of continuous profit lock-in, there is only one regime which is similar to the transition regime of the periodic case.

\subsubsection{Pricing}

The terminal payoffs we naturally want to consider are
\begin{itemize}
\item $C_n = G_n X_n$, the CPPI strategy;
\item $\max (C_n,G_n) = G_n \max (X_n,1)$, the Guaranteed CPPI strategy;
\item $\left( C_n - K G_n \right)^+ = G_n \left( X_n - K \right)^+$, a Call on CPPI with strike in percentage of the final guarantee;
\item $\left( K G_n - C_n \right)^+ = G_n \left( K - X_n \right)^+$, a Put on CPPI with strike in percentage of the final guarantee;
\item $\mathbbm{1}_{ C_n < K G_n }$, a digital Put with strike in percentage of the final guarantee ($K=1$ gives the defeasance probability);
\item $\left( C_n - K \right)^+ $, a Call on CPPI with strike in monetary unit;
\item $\left( K - C_n \right)^+ $, a Put on CPPI with strike in monetary unit;
\item $\mathbbm{1}_{ C_n < K G_n }$, a digital Put with strike in monetary unit.
\end{itemize}
With the exception of options with strike in monetary unit, these payoffs are of the form $G_n P(X_n)$ or $P(X_n)$. For a payoff which depends directly on the monetary value $C_T$ the computation is different although the basic building blocks remain the same (see section \ref{appendixpayoff}).

All computations throughout this section will be made with the hypothesis that there are no coupons. We will see at the end of the section that there is no difficulty in including them.

With profit lock-in, the single variable $C_i$ is no longer Markovian. One should introduce in addition the guarantee to construct a two-dimensional Markov model. However, the process $\frac{C_i}{H_i}$ (where the threshold $H_i$ is proportional to the guarantee $G_i$) is Markovian: the evolution of this quantity over one lock-in period depends only on its value at the beginning of the period, and periods are independent of each other.
We introduce therefore two pricing kernels:
\begin{eqnarray*}
\mathcal K^{(0)}(x,y) &=& \mathbb{E} \left[ \delta( X_n - y ) \, \vert \, X_0 = x \right]
\\
\mathcal K^{(1)}(x,y) &=& \mathbb{E} \left[ \frac{G_n}{G_0} \delta( X_n - y ) \, \big\vert \, X_0 = x \right]
\end{eqnarray*}
(Without profit lock-in, both kernels are equal, this is the kernel $\mathcal K$ of section \ref{section-multiperiod}.) $\mathcal K^{(0)}$ is used to compute the defeasance probability or digital options:
\begin{equation}
\mathbb{E}\!\left[P(X_n)\, \vert \, X_0\right] = \left(\mathcal K^{(0)}P\right)(X_0)
\label{eq-LI-digital}
\end{equation}
whereas $\mathcal K^{(1)}$ is used to compute payoffs of the form $G_n P(X_n)$:
\begin{equation}
\mathbb{E} \!\left[G_n P(X_n)\, \big\vert \, X_0 \right] = G_0 \mathbb{E}\!\left[ \frac{G_n}{G_0} P(X_n) \, \big\vert \, X_0 \right] = G_0 \left(\mathcal K^{(1)}P\right)(X_0)
\label{eq-LI-payoff}
\end{equation}

To compute the pricing kernels, the first step is to compute the evolution operator between two lock-in dates. We consider a process which ends just before the lock-in at date $t_{I+1}$, that we denote by $t_{I+1}^-$. From section \ref{section-multiperiod}, the evolution operator on this period is given by
\begin{equation}
\mathcal K_{I,I+1^-} = \mathcal M_{l(I)} \mathcal M_{l(I)+1} \cdots \mathcal M_{l(I+1)-1}
\label{eq-betweenLI}
\end{equation}
Once this operator is computed, the lock-in has to be incorporated. It is given by a map which acts as
\begin{equation*}
X_{I+1}^- \longrightarrow X_{I+1} = \frac{X_{I+1}^-}{f_{I+1}(X_I,X_{I+1}^-)}
\end{equation*}
Compounding this map with the operator without profit lock-in gives the transition kernel on the lock-in period, including lock-in:
\begin{equation*}
\mathcal V^{(0)}_{I}(x,y) = \mathbb{E} \left[ \, \delta( X_{I+1} - y ) \, \vert \, X_I = x \right] = \int \mathrm{d}z \, \mathcal K_{I,I+1^-}(x,z) \, \delta\!\left(y-\frac{z}{f_{I+1}(x,z)}\right)
\end{equation*}
When the multiplying factor of the guarantee is included, it gives the second kernel
\begin{equation*}
\mathcal V^{(1)}_{I}(x,y) = \mathbb{E} \left[ \frac{G_{I+1}}{G_I} \, \delta( X_{I+1} - y ) \, \big\vert \, X_I = x \right] = \int \mathrm{d}z \, \mathcal K_{I,I+1^-}(x,z) \, \delta\!\left(y-\frac{z}{f_{I+1}(x,z)}\right) f_{I+1}(x,z)
\end{equation*}
As one can see in Figure~\ref{factorf}, there might be, depending on the value of $X_{I+1}$, zero, one, two or an infinity of values of $X^-_{I+1}$ corresponding to different states of the portfolio before rebalancing.

As the processes for $X_I$ on different periods are independent, the total pricing kernels are given by the operator products
\begin{eqnarray*}
\mathcal K^{(0)} &=& \prod_I \mathcal V^{(0)}_{I}
\\
\mathcal K^{(1)} &=& \prod_I \mathcal V^{(1)}_{I}
\end{eqnarray*}
They can be used in formulas \eqref{eq-LI-digital} and \eqref{eq-LI-payoff} to compute the expected value of the quantity considered.

To account for coupons, the only modification to make is that in the definition of the operator $\mathcal K_{I,I+1^-}$, $\mathcal M_i$ has to be replaced by $\widetilde{\mathcal M}_i$ the evolution operator in the presence of coupons. The pricing then goes along the same lines as in the coupon case, making use of the same semi-direct product.

\subsubsection{Options with strike in absolute amount}
\label{appendixpayoff}

With a strike in absolute amount, or equivalently in percentage of the initial guarantee, the computation scheme is different.

Here we write a transition matrix directly for the CPPI index value $C_i$. Technically we consider a deterministic rescaling which defines the variable
\begin{equation*}
Z_i = \frac{C_i}{G_0} e^{\int_{t_i}^T \left( r+s_H \right) \, \mathrm{d}t} = \frac{C_i G_i}{H_i G_0} = X_i \frac{G_i}{G_0}
\end{equation*}
We want to compute the transition operator for this variable
\begin{equation*}
\mathcal U(x,y) = \mathbb{E}\!\left[ \delta \left( Z_n - y \right) \, \vert \, Z_0 = x \right]
\end{equation*}
This transition operator gives the probability distribution at the maturity of the CPPI and allows the pricing of payoff $P(C_n)$ as
\begin{equation*}
\mathbb{E} \left[P(C_n) \, \vert \, Z_0 \right] = (\mathcal U P)(Z_0)
\end{equation*}

The transition operator can be computed by recursion. Introducing the evolution operator from lock-in $I$ to maturity
\begin{equation*}
\mathcal U_I(x,y) = \mathbb{E}\!\left[ \delta \left( \frac{Z_n}{G_I} - y \right) \, \big\vert \, \frac{Z_I}{G_I} = X_I = x \right]
\end{equation*}
the basic step is the recursion formula
\begin{equation*}
\mathcal U_{I}(x,y) = \int \mathrm{d}z \, \mathcal K_{I,I+1^-}(x,z) \, \mathcal U_{I+1}\!\left( \frac{z}{f_{I+1}(x,z)}, \frac{y}{f_{I+1}(x,z)} \right) \frac{1}{f_{I+1}(x,z)}
\end{equation*}
where $\mathcal K_{I,I+1^-}$ is given by equation \eqref{eq-betweenLI}.
This equation states that the evolution after a lock-in is the same evolution as without any lock-in but for a rescaled variable.
The cumulative operator
\begin{equation*}
\mathcal Q_{I}(x,y) = \int_{-\infty}^{y} \mathrm{d}z \, \mathcal U_{I}(x,z)
\end{equation*}
has a simpler recursion formula, more adapted to numerical interpolation:
\begin{equation*}
\mathcal Q_{I}(x,y) = \int \mathrm{d}z \, \mathcal K_{I,I+1^-}(x,z) \, \mathcal Q_{I+1}\!\left( \frac{z}{f_{I+1}(x,z)} , \frac{y}{f_{I+1}(x,z)} \right)
\end{equation*}
Using this equation backward from the CPPI maturity, the full transition matrix $\mathcal U = \mathcal U_0$ can be computed and gives the distribution of the final CPPI index value, the price of the CPPI strategy or any option with absolute strike. Note that this computation scheme is no longer a Markov chain: it can not be reduced to a point wise rescaling followed by a matrix multiplication.

\subsection{Conditional rebalancing}
\label{section-conditionalrebalancing}

In order to minimize transaction costs and to simplify the management of the portfolio, one might choose not to rebalance the position if it is close enough to the previous one. For instance, one starts at time $t_0$ at a given level $X_0$ with the risky asset weighting level $W_0 = w(X_0)$. The CPPI will be rebalance on $t_1$ only if (the theoretical weighting) $w(X_1)$ is such that $w(X_1) \notin [(1-\beta)W_0, (1+\beta)W_0 ]$ for a given $\beta > 0$. If $w(X_1)$ lies within the interval, the risky asset weighting for the next period is left at $W_0$. This makes $W_1$ a function of both $X_0$ and $X_1$. On the following rebalancing date, $W_2$ would be a function of $X_0$, $X_1$ and $X_2$ and so on... To preserve the Markov property, we note that in this setup $W_i$ is a function of $W_{i-1}$ and $X_i$ only: we have to follow the diffusion of both $X$ and $W$, turning our problem 2 dimensional. The dynamics of $W$ is given by
\begin{equation*}
W_{i+1} = W_i \mathbbm{1}_{ \left|\frac{w(X_{i+1})}{W_i} - 1\right| \leq \beta} + w(X_{i+1}) \mathbbm{1}_{ \left|\frac{w(X_{i+1})}{W_i} - 1\right| > \beta}
\end{equation*}

Instead of keeping track of $W$, it is more convenient to change variable and set
\begin{equation*}
W_i = \frac{w(X_i)}{1 + Y_i}
\end{equation*}
The new variable $Y_i$ measures the the difference between the exposure $W_i$ at time $t_i$ and its "theoretical value" $w(X_i)$. If one defines
\begin{equation*}
 \alpha_{i+1} = \frac{w(X_{i+1})}{W_i} - 1 =  \frac{w(X_{i+1})}{w(X_i)} ( 1 + Y_i ) - 1
\end{equation*}
the dynamics of $Y$ is given by
\begin{equation*}
Y_{i+1} = \alpha_{i+1} \mathbbm{1}_{\alpha_{i+1} \in [-\beta, \beta]}, \quad Y_0 = 0
\end{equation*}

The main advantage of considering this new variable over direct propagation of $W$ is that for typical rebalancing rules, $\beta$ is equal to a few percents and hence $Y$ is constrained into a small interval which we can approximate with a small number of points in the $Y$-direction keeping the two dimensional problem tractable on a computer. The equation driving the dynamics of the CPPI over one period (see \eqref{eq-SF-oneperiod} and following) still hold if one substitutes $w(X_i)$ with $W_i$.

\subsection{Open-ended CPPI}

An open-ended CPPI has no fixed maturity: the investor can exit at any time. It is therefore an American-style product. In addition, it has continuous profit lock-in: the guarantee is set to some proportion $p$ of the highest value attained. As there is no fixed maturity, the threshold cannot be discounted, it is equal to the guarantee.

As there the threshold is flat between rebalancing, the threshold spread is set over each period to the opposite of the interest rate $s_H = -r$ such that $s_H+r = 0$. The continuous lock-in is handled as explained in section \ref{section-LI}: on each period the evolution matrix $\mathcal M_i$ is used to compute the two pricing kernels $\mathcal V_i^{(0)}$ and $\mathcal V_i^{(1)}$. The American feature is taken into account by propagating backwards the payoff vector, taking the maximum with the exercise payoff vector at each step. At date 0 this gives the expected value of the American product conditional to the starting value of the CPPI index.

If there is no terminal maturity, we force some long range maturity by hand, so that the process can be bootstrapped. If the put price diverges when this terminal maturity goes to infinity, the fair price of the product in the model considered is infinite, which probably means that the fees are not high enough to compensate for the gap risk.

\section{Numerical aspects}
\label{section-numerical}

In order to make practical computation, one discretizes the state space by a finite space (the grid) $g = \{ g_1, \dots, g_N \}$. We also introduce an "separation" grid: $g' = \{ g'_1, \dots, g'_N \}$ such that $g_{k-1} \leq g'_k \leq g_k $.
  Linear operators are approximated by matrices on this grid. Non linear features such as profit lock-ins are obtained by point wise operations on matrix element. Hereafter, we will denote the discretized operator corresponding to an operator $\mathcal O$ by $\overline{\mathcal O}$. In our implementation, the discretization scheme goes as follows: one replaces the probability density $\mathbb{P} \left[ X_{i+1} = y \, \big\vert \, X_i = x \right]$ by a matrix $\overline P^{(i,i+1)}$ whose elements $\overline P^{(i,i+1)}_{kl}$ are given by $(1 \leq k,l \leq N)$
\begin{equation*}
\overline P^{(i,i+1)}_{kl} = \int_{g'_{l-1}}^{g'_l} dy \ \mathbb{P} \left[ X_{i+1} = y \, \big\vert \, X_i = g_k \right]
\end{equation*}
where  The Markov property
\begin{equation*}
\mathbb{P} \left[ X_{i+2} = z \, \big\vert \, X_{i} = x \right] = \int \mathrm{d}y \, \mathbb{P} \left[ X_{i+2} = z \, \big\vert \, X_{i+1} = y \right] \ \mathbb{P} \left[ X_{i+1} = y \, \big\vert \, X_{i} = x \right]
\end{equation*}
is then replaced by the relation:
\begin{equation*}
\overline P^{(i,i+2)}_{km} = \sum_{l=1}^N \overline P^{(i+1,i+2)}_{lm} \overline P^{(i,i+1)}_{kl}
\end{equation*}
This discretization scheme amounts to considering that, starting form a point $X_i = x$, if we arrive at a point $X_{i+1}$ such that $g'_k \leq X_{i+1} \leq g'_{k+1}$, the next step will start at $X_{i+1} = g_k$.


\subsection{Grids}

We take a fixed grid for the whole life of the product. The pricing could be improved by changing the grid extension as the maturity grows up, but it does not appear to be so important as long as the variance of individual steps is controlled.

The upper bound of the grid is computed using the lognormal, continuous approximation. The process $C-H$ is supposed to be log-normally distributed with volatility $w_\mathrm{max} \sigma$ where $w_\mathrm{max}$ is the maximum exposure, equal to the multiplier if no specific maximum exposure is set. $\sigma$ is the At-The-Money volatility of the underlying asset with the same maturity. Inverting the cumulative law of this lognormal distribution for some value $1-\epsilon$ close to 1 provides an upper bound for the grid. Because of the leverage, this bound can be very high for a long maturity.

The lower bound of the grid is constructed from the upper bound. If the risky asset goes to zero or close to zero, the CPPI index goes from $C$ to $(1-w)C$. Therefore the lower bound is chosen as $(1-w_\mathrm{max})$ times the upper bound. It is generally a deeply negative value.

Between these two points, the grid is densified around the threshold, where the behavior of the index changes and where the distribution can be sharp. If there is some cushion limit which stops investment in the risky asset (see section \ref{section-cushion-limit} for a precise definition), the grid is exponential between the threshold and the cushion limit: in this case there is only a drift motion in this region and a grid too large would introduce diffusion. We detail below how this diffusion is prevented using negative probabilities.

The "separation" grid is introduced to separate points of the first grid and defines a partition of the real axis. When computing a transition probability, the probability to be at a given grid point is defined as the probability to lie inside the interval around this point, computed by difference from the cumulative function. This ensures that the sum of all probabilities is always 1.

\subsection{Matrix-Matrix vs Matrix-Vector multiplications}

When there is no explicit time dependence, the computation of the price amounts to propagating a vector through the $n$\textsuperscript{th} of a matrix. Numerically, there are two ways of doing the computation. The first solution is to take the $n$\textsuperscript{th} power of the matrix M by fast exponentiation \cite{albanese2006oma}: $M^{2^p}$ are computed recursively for growing $p$ by $M^{2^p} = M^{2^{p-1}} M^{2^{p-1}}$. The power $n$ is decomposed in basis 2 as
\begin{equation*}
n = \sum_i 2^{p_i}
\end{equation*}
and $M^n$ is obtained by multiplying the corresponding $M^{2^{p_i}}$:
\begin{equation*}
M^n = \prod_i M^{2^{p_i}}
\end{equation*}
For a discretization on a grid of size $N$, matrix multiplication has complexity $O(N^3)$. The number of multiplication to perform is $O(\ln n)$. Thus this solution has complexity $O( \ln(n) N^3)$. The other solution is to sequentially apply $n$ times the operator $M$ to the vector $P$. The matrix vector multiplication being of complexity $O(N^2)$ this gives a total complexity in $O(n N^2)$. Depending of $N$, $n$ and of the architecture used to perform the linear algebra, one will choose the fastest one for his particular needs.

\subsection{Mean and Variance control}

The discretization of the transition kernel introduce a bias in the mean value of the index $C$. To prevent this and get the correct forward price, the probability distribution constructed is slightly shifted on the grid in the opposite direction, such that the mean value computed on the grid equals the theoretical mean value, conditionally to the starting point.

Similarly, the variance of the process conditionally to the starting point is also adjusted. If the variance computed on the grid is too high, the probabilities above and below the mean value are slightly displaced closer to the mean value, in a way which preserves the mean. The converse is done when the variance computed on the grid is lower than the theoretical variance.

Finally, we get probability distribution conditional to the starting point which gives the correct mean and variance computed on the grid. This appears to be crucial for convergence when the number of rebalancing dates grows up. If the mean is not controlled, the expected value of the CPPI strategy can diverge. If the variance is not controlled at each step, the total variance is not correct and a reasonable estimate of the gap probability can require a large number of points on the grid, which slows down the computation in $N^2$.

\subsection{Convective limit}

When the threshold does not grow at the same rate as the risk-free rate, the process $X$ has some drift. In region where the pure diffusive part is small (\emph{i.e.} near the threshold), this drift might have undesired effects: unless the drift over one rebalancing period exactly corresponds to an integer number of steps in the grid, the discretization scheme will introduce some spurious diffusion. This enlarges the distribution in this important part of the grid where the threshold lives and can lead to very inaccurate results for the gap probability and mean value. This phenomenon is more easily identified when the CPPI includes a cushion limit in which case the CPPI structure is fully monetized before reaching the threshold.

An efficient way to suppress this problem is to allow negative probabilities such that the variance can be set to zero even when the process must be between two points of the grid. Let consider that starting from the grid point $g_i$, the process must drift to $g_{i-1} < x < g_i$. If no negative probabilities are allowed, the solution which minimizes the variance while preserving the mean value is to take probabilities
\begin{eqnarray*}
p_{i-1} &=& \frac{x - g_i}{g_{i-1} - g_i} \\
p_{i} &=& 1-p_{i-1}
\end{eqnarray*}
The mean value is $x$ but the variance is not zero. We consider instead the following "transition probabilities":
\begin{eqnarray*}
p_{i-3} & = & \alpha q \\
p_{i-2} & = & -q \\
p_{i-1} & = & p \\
p_{i}   & = & 1 - p + (2-\alpha) q \\
p_{i+1} & = & -q
\end{eqnarray*}
where $p$ and $q$ are positive and $\alpha$ is a numerical constant. Taking a non vanishing $\alpha$ helps stabilizing the process. We set it to $\alpha = 0.1$ as this value gives good results for both stability of the process and convergence of the gap estimators. $p$ and $q$ are chosen so that the mean value is $x$ and the variance vanishes. These conditions translates into a linear system of two equations  in $p$ and $q$ which is easily solved. We bound the negative probability $q$ to $0.05$ to avoid explosions of negative probabilities in some parts of the grid.

This drift scheme appears to be impressively better than the naive one. The accuracy obtained for a grid of 2000 points can be reached with a few hundreds of points. Drawbacks of negative gap probabilities or gap mean occur only for very sparse grids. Further work would be necessary to be able to control a priori the stability of this trick.

\subsection{Semi-direct product in matrix form}

The semi-direct product we defined in the section 4.3 is easily handled with matrices. If $N$ is the size of the discretization, to a pair $\left( \mathcal O, f \right)$ where $\mathcal O$ is an operator and $f$ is a function, we associate the discretization given by the $(N+1) \times (N+1)$ block matrix
\begin{equation*}
\left( \begin{array}{ccc|c}
 &               & &   \\
 & \overline{\mathcal O}    & & \overline{f} \\
 &               & &   \\ \hline
 &    0          & & 1
 \end{array} \right)
\end{equation*}
The product $\left( \mathcal O_1, f_1 \right) \left( \mathcal O_2, f_2 \right)$ of two such matrices is given by
\begin{equation*}
\left( \begin{array}{ccc|c}
 &               & &   \\
 & \overline{\mathcal O_1} \ \overline{\mathcal O_2}    & & \overline{\mathcal O_1} \ \overline{f_2} + \overline{f_1} \\
 &               & &   \\ \hline
 &    0          & & 1
 \end{array} \right)
\end{equation*}
which is precisely the same algebra as the semi-direct product we previously defined.

The action of a pair $\left( \mathcal O, f \right) $ on a function $g$ is now represented by the standard product of matrices
\begin{equation*}
\left( \begin{array}{ccc|c}
 &               & &   \\
 & \overline{\mathcal O}   & & \overline f \\
 &               & &   \\ \hline
 &    0          & & 1
 \end{array} \right)
\cdot
\left( \begin{array}{c}
   \\
\overline g \\
   \\ \hline
 1
 \end{array} \right)
= \left( \begin{array}{c}
   \\
\overline{\mathcal O} \cdot \overline g + \overline f\\
   \\ \hline
 1
 \end{array} \right)
\end{equation*}

\subsection{Piecewise constant approximation}

Quantities involved in the computation of a single period transition matrix depend on the precise features of the CPPI. There is at least the variance of the risky asset over the considered period. When spreads over the risk-free rate are applied, the value of the zero-coupon spread over the period can change, if the length of the period changes a bit, depending on the day count basis, or if the spread is not expressed in continuous mode and the interest rate is not constant. Moreover when there are coupons or continuous profit lock-in, the zero-coupon value between the rebalancing date and the maturity of the CPPI enters also the computation. In these cases, an exact computation would require computing all single period matrices. The computation can be fastened if such quantities are approximated by piecewise constant functions. Technically, we choose a tolerance for the variation of variables and we compute only one matrices for periods with parameters which do not differ by values larger than this tolerance. Mean values for the parameters are computing in the following way. The variance of the risky asset is the arithmetic mean of the variance over single periods whereas for zero coupons it is the geometric mean which is used, such that their product is equal to the real zero-coupon over the sum of periods considered. The numerical impact of this approximation is studied in section \ref{section-convergence}.

\section{Numerical convergence}
\label{section-convergence}

In order to analyze the impact of the grid size, we set up two benchmark CPPIs and study the price of a put option with a strike equal to the guarantee. Both CPPI start on November 12, 2008 and end on November 7, 2018, their multiplier is $4$, their nominal is 1 000 000 and are rebalanced every week. On the start date, the risky underlying is quoted at $3207$.
\begin{itemize}
	\item {\bf Vanilla CPPI}. The rate curve is flat with a constant rate of 5\%, the volatility of the risky underlying is 50\%. Threshold is natural and there are no fees.
	\item {\bf Non-Vanilla CPPI}. Volatility is still 50\% but the rates are time dependent. The threshold is linear with an initial value of 60\% of the nominal. There are a cushion limit\footnote{See section \ref{section-cushion-limit}.} of 5\% and proportional fees of 50bp.
\end{itemize}
In the following, we consider a European put option with a strike given by the guaranteed nominal (there is no profit lock-in). The computation date is chosen to be November 16, 2008 and the spot of the risky underlying on that date is $3190$.

The computations (both Markov and Monte Carlo methods) were performed on a PC equipped with an Intel Core2 6700 at 2.66 GHz and 3.25 GB of memory. For the Markov method, we used Intel's Math Kernel Library to perform matrix-matrix and matrix-vector multiplications. The source codes are not heavily optimized and hence computing times are only relevant for comparison purposes. In the case of Markov, the times we indicate correspond to only one pricing which is sufficient for getting the price, the delta and the gamma of the option. In order to get the vega two additional pricing are performed (but only one more could be enough).

\subsection{Vanilla CPPI}

\newlength{\pourcentrer}
\settowidth{\pourcentrer}{\begin{tabular}{lr}
Put Price & 170.5530 \\
Put Delta & 0.2177   \\
Put Vega  & 68.2553   \\
Gap Proportion (\%) & 9.7989
\end{tabular}}
\addtolength{\pourcentrer}{-\textwidth}

In the case of the vanilla CPPI, we can compute analytically the price and the sensibilities of the option using the formulas of appendix \ref{appendix-closed-formula}. The results obtained are (these numbers have to be compared to the CPPI initial value of 1 000 000):\\
\hspace*{-.5\pourcentrer}\begin{tabular}{lr}
&\\
Put Price & 170.5530 \\
Put Delta & 0.2177   \\
Put Vega  & 68.2553   \\
Gap Proportion (\%) & 9.7989 \\
&
\end{tabular}

In table \ref{conv-vanilla}, we give the relative errors with respect to the analytical results and the computation time, depending on the grid size. Even with a small number of points, there are almost no pricing errors: this is due to the fact that we (try to) force the distributions on each rebalancing period to have the correct mean, variance, conditional (on defeasance on that rebalancing date) mean and conditional variance: the price of the option and the gap proportion of the CPPI are then automatically exact.
\begin{table}[h]
\centering
\begin{tabular}{|c|c|c|c|c|c|}
\hline
Grid Size  & Price & Delta & Vega & Gap Prop. & Time (s) \\
\hline
50	 & $-3.09\times 10^{-8}$ &	$-6.57\times 10^{-3}$ &	$-6.34\times 10^{-6}$ &	$-3.32\times 10^{-10}$ &	0.06  \\
100	 & $-9.02\times 10^{-8}$ &	$-6.89\times 10^{-5}$ &	$-6.52\times 10^{-6}$ &	$-3.32\times 10^{-10}$ &	0.07  \\
250	 & $-4.50\times 10^{-8}$ &	$-2.68\times 10^{-6}$ &	$-6.20\times 10^{-6}$ &	$-3.32\times 10^{-10}$ &	0.10  \\
500	 & $-5.58\times 10^{-8}$ &	$-6.35\times 10^{-7}$ &	$-6.42\times 10^{-6}$ &	$-3.32\times 10^{-10}$ &	0.24  \\
750	 & $-5.38\times 10^{-8}$ &	$-3.57\times 10^{-7}$ &	$-6.41\times 10^{-6}$ &	$-3.33\times 10^{-10}$ &	0.59  \\
1000 & $-5.68\times 10^{-8}$ &	$-2.55\times 10^{-7}$ &	$-6.42\times 10^{-6}$ &	$-3.33\times 10^{-10}$ &	1.49  \\
1500 & $-5.71\times 10^{-8}$ &	$-1.66\times 10^{-7}$ &	$-6.42\times 10^{-6}$ &	$-3.33\times 10^{-10}$ &	3.80  \\
2000 & $-5.67\times 10^{-8}$ &	$-1.31\times 10^{-7}$ &	$-6.42\times 10^{-6}$ &	$-3.32\times 10^{-10}$ &	7.33  \\
2500 & $-5.69\times 10^{-8}$ &	$-1.09\times 10^{-7}$ &	$-6.43\times 10^{-6}$ &	$-3.32\times 10^{-10}$ &	12.41 \\
\hline
\end{tabular}
\caption{\emph{Relative errors for the put on the vanilla CPPI.}}
\label{conv-vanilla}
\end{table}
The vega being computed by finite difference (with a fixed bump) on the price has almost no dependance on the grid size. The delta is computed by finite difference on the grid and hence slightly depends on the grid size by a third order effect. The non-vanilla case will give more realistic errors and provide a better estimate of the necessary grid size and computation time. This test case was intended to check the algorithm in a case where closed formulas were known.

\subsection{Monte-Carlo pricing}

\begin{figure}[htbp]
\centering
\includegraphics[width=.75\textwidth]{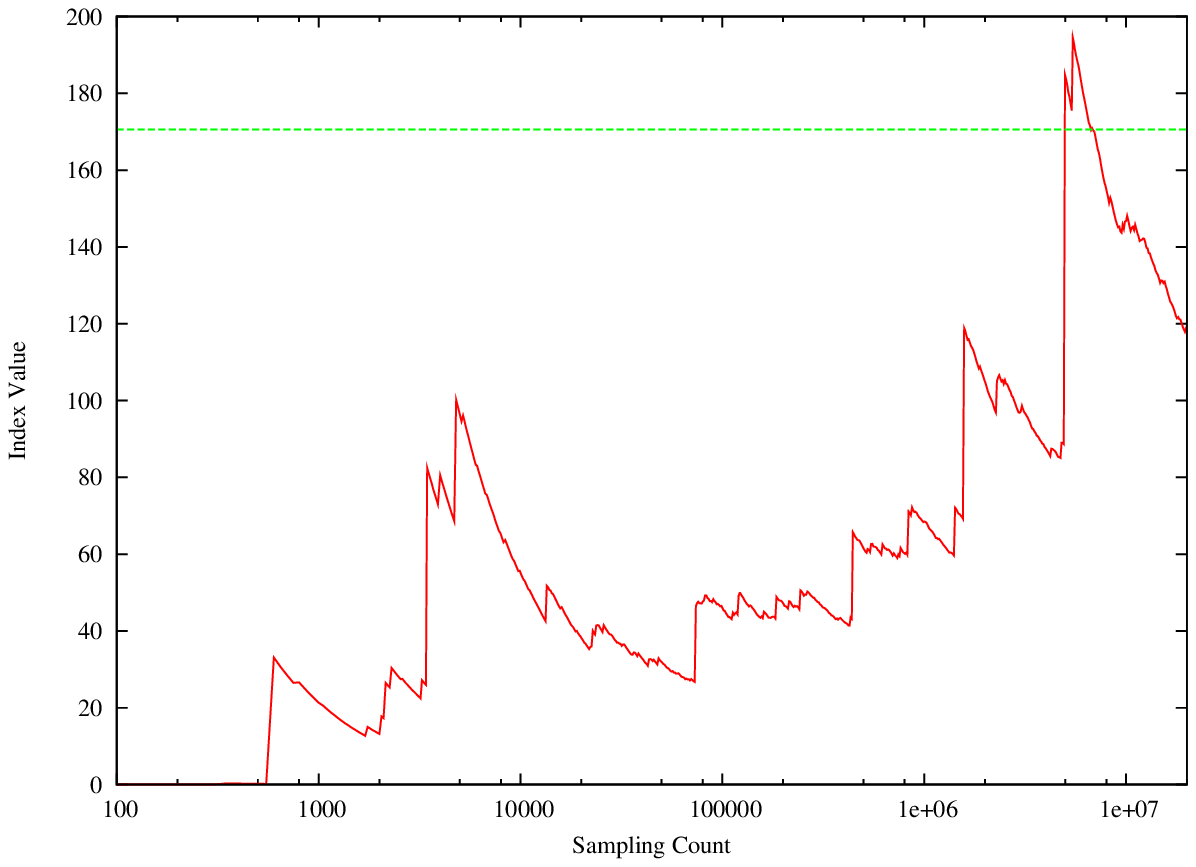}
\includegraphics[width=.75\textwidth]{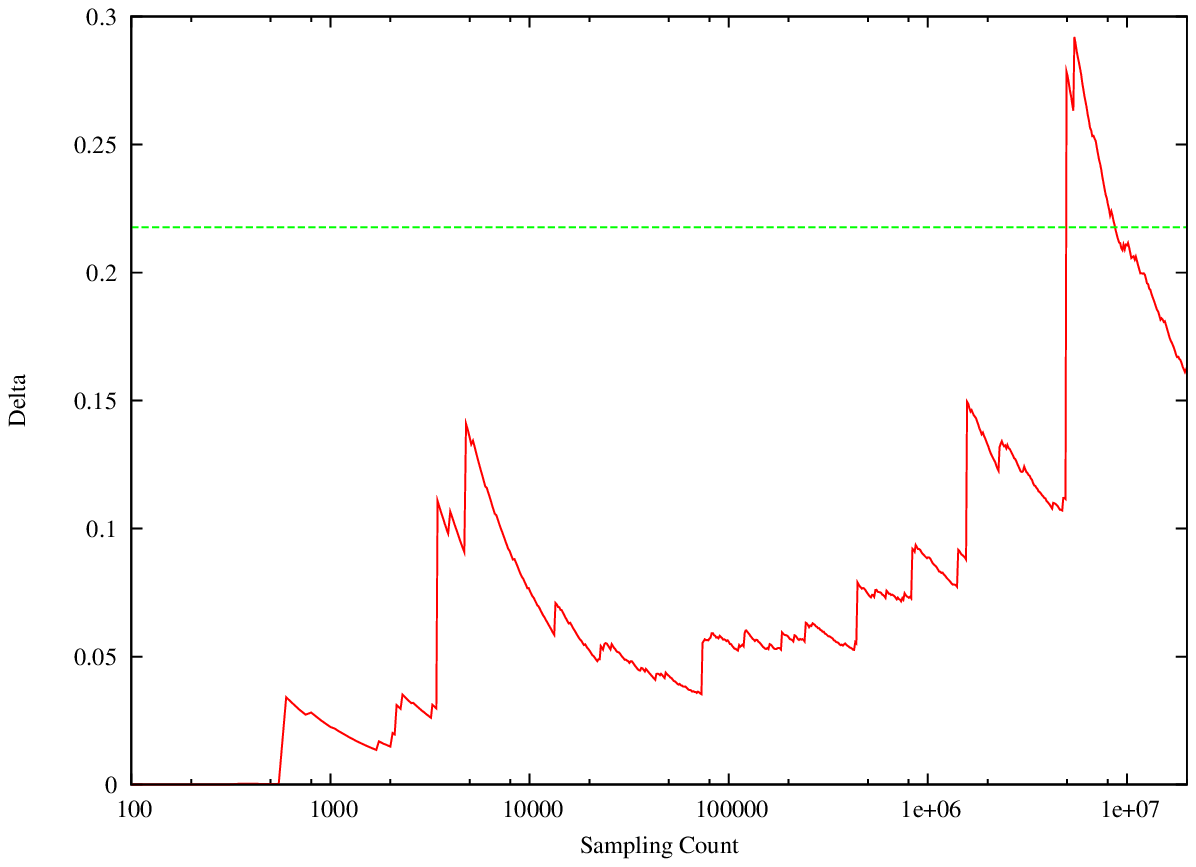}
\includegraphics[width=.75\textwidth]{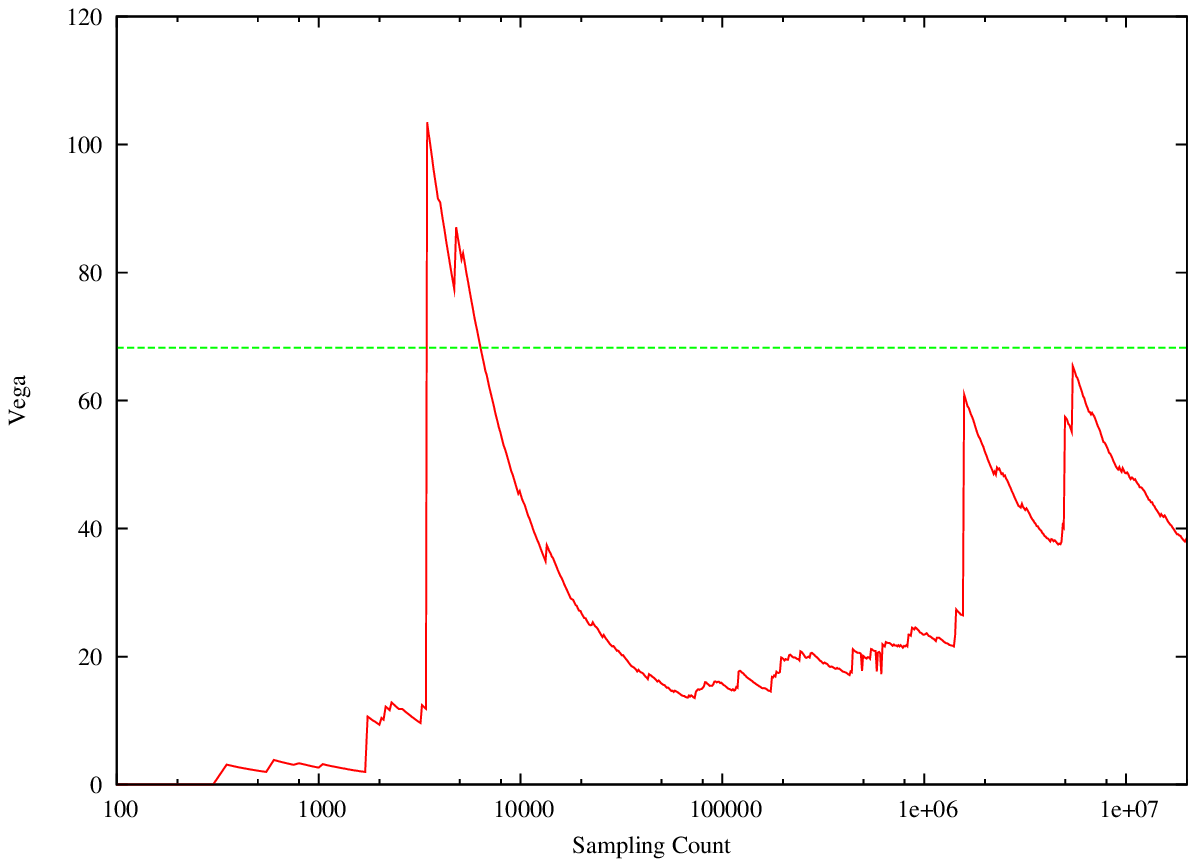}
\caption{\emph{Convergence of the Monte Carlo method for price, delta and vega in the vanilla case, as functions of the number of paths. In each graph, the dashed line indicates the corresponding exact value obtained with analytical computation.}}
\label{fig-vanilla-mc}
\end{figure}

We compare our results to a standard Monte-Carlo pricing run on the same computer. We take a Sobol random generator and transform the uniform laws into normal laws through a Moro algorithm. In the continuous rebalancing limit, the CPPI terminal value depends only on the underlying spot at maturity. So we use a Brownian bridge path generation which takes advantage of this property and improves convergence. Figure \ref{fig-vanilla-mc} shows the convergence graph up to 20 000 000 paths. In 18 hours, the price and greeks have not converged\footnote{Other variance reduction techniques could have been used. With importance sampling where more weight would be given to tail events. The knowledge of the closed formula in the vanilla case can also be used to implement a control-variate for non-vanilla cases. However jumps in the price due to extreme events would not necessarily occur at the same time for both products: this could produce a very jumpy estimate of the covariance. It is not clear if the convergence would be better.}.

\subsection{Grid size}

To study the influence of the grid size on the outputs we consider now the non-vanilla case. Table \ref{conv-nonvanilla} shows the put price, delta, gamma, and defeasance probability depending on the grid size. The precision is correct from 100 points, and quite good from 500 points. Note that the goal of this example is to study the numerical precision of the pricing model: as the vega is very high compared to the price, even with a grid size of 100 the numerical errors are negligible in front of model risk. 

\begin{table}[htbp]
\centering
\begin{tabular}{|c|r|r|r|r|}
\hline
Grid Size  & \multicolumn{1}{c|}{Price} & \multicolumn{1}{c|}{Delta} & \multicolumn{1}{c|}{Vega} & \multicolumn{1}{c|}{Gap Prop.} \\
\hline
50	 &  2028.6075 &	3.3718 & -48.0019 &	$-6.7264\times 10^{-3}$ \\
100	 &  129.7479  & 0.1902 & 52.9596  &	$5.2489\times 10^{-4}$  \\
250	 &  133.4823  & 0.1941 & 54.4288  &	$5.8667\times 10^{-4}$  \\
500	 &  138.3577  & 0.2016 & 56.3636  &	$5.9074\times 10^{-4}$  \\
750	 &  139.3600  & 0.2034 & 56.7991  &	$5.7116\times 10^{-4}$  \\
1000 &	139.7006  &	0.2040 & 56.9460  &	$5.6653\times 10^{-4}$  \\
1500 &	139.9571  &	0.2044 & 57.0554  &	$5.6465\times 10^{-4}$  \\
2000 &	140.0305  &	0.2045 & 57.0868  &	$5.6360\times 10^{-4}$  \\
2500 &	140.0549  &	0.2046 & 57.0972  &	$5.6350\times 10^{-4}$  \\
\hline
\end{tabular}
\caption{\emph{Convergence results for the put on the non-vanilla CPPI.}}
\label{conv-nonvanilla}
\end{table}

\subsection{Piecewise constant approximation}

In order to speed up the pricing, the interest and threshold rates are approximated by piecewise constant functions as well as variance of the underlying between two rebalancing dates. The approximation is controlled by a tolerance number: when the rate has moved by more than this number a new constant is used. Constant rates are calibrated so that zero-coupons are exact on the period. This reduces in a large proportion the number of matrices to be generated. Computation times depending on the tolerance level and the grid size are presented in table \ref{conv-nv-time}. Relative errors with respect to the most accurate case (exact rates, 2500 points) are presented in tables \ref{conv-nv-price}, \ref{conv-nv-delta}, \ref{conv-nv-vega} and \ref{conv-nv-gp}. A good precision can be reached in about 1s.

\newcommand{\withrelerr}[2]{$\displaystyle\mathop{#1}_{\tiny #2}$}
\begin{table}[p]
\centering
\begin{tabular}{|c|ccccc|}
\hline
 & \multicolumn{1}{c}{0\,\%}&\multicolumn{1}{c}{5\,\%} & \multicolumn{1}{c}{20\,\%} & \multicolumn{1}{c}{50\,\%} & \multicolumn{1}{c|}{100\,\%} \\
\hline
50	 & \withrelerr{0.76}{1.35\times 10^{+1}}  &  \withrelerr{0.13}{1.35\times 10^{+1}}  & \withrelerr{0.10}{1.36\times 10^{+1}}  &	 \withrelerr{0.08}{1.37\times 10^{+1}} & \withrelerr{0.07}{1.40\times 10^{+1}} \\

100  & \withrelerr{1.61}{-7.36\times 10^{-2}}  &\withrelerr{0.19}{-7.34\times 10^{-2}} &	\withrelerr{0.12}{-7.20\times 10^{-2}}  &	 \withrelerr{0.09}{-6.07\times 10^{-2}}  &	 \withrelerr{0.08}{-3.22\times 10^{-2}} \\

250  & \withrelerr{5.26}{-4.69\times 10^{-2}}   &	\withrelerr{0.44}{-4.67\times 10^{-2}}  &	\withrelerr{0.21}{-4.53\times 10^{-2}}  &	 \withrelerr{0.14}{-3.43\times 10^{-2}}  &	 \withrelerr{0.12}{-5.94\times 10^{-3}} \\

500	 & \withrelerr{15.33}{-1.21\times 10^{-2}}   &	\withrelerr{1.15}{-1.19\times 10^{-2}}  &	\withrelerr{0.51}{-1.05\times 10^{-2}}  &	 \withrelerr{0.31}{7.52\times 10^{-4}}  &	 \withrelerr{0.28}{3.12\times 10^{-2}} \\

750	 & \withrelerr{30.53}{-4.96\times 10^{-3}}   &	\withrelerr{2.33}{-4.77\times 10^{-3}}  &	\withrelerr{1.07}{-3.30\times 10^{-3}}  &	 \withrelerr{0.73}{8.45\times 10^{-3}}  &	 \withrelerr{0.66}{3.83\times 10^{-2}} \\

1000 & \withrelerr{50.62}{-2.53\times 10^{-3}}  &	\withrelerr{4.38}{-2.34\times 10^{-3}}  &	\withrelerr{2.32}{-8.90\times 10^{-4}}  &	 \withrelerr{1.74}{1.07\times 10^{-2}}  &	 \withrelerr{1.64}{4.00\times 10^{-2}} \\

1500 & \withrelerr{109.31}{-6.99\times 10^{-4}} &	\withrelerr{10.20}{-5.11\times 10^{-4}} &	\withrelerr{5.69}{9.28\times 10^{-4}}  &	 \withrelerr{4.41}{1.24\times 10^{-2}}  &	 \withrelerr{4.17}{4.15\times 10^{-2}} \\

2000 & \withrelerr{173.11}{-1.75\times 10^{-4}} &	\withrelerr{17.70}{1.47\times 10^{-5}} &	\withrelerr{10.46}{1.46\times 10^{-3}} &	 \withrelerr{8.46}{1.29\times 10^{-2}}  &	 \withrelerr{8.17}{4.22\times 10^{-2}} \\

2500 & \withrelerr{266.14}{-} &	\withrelerr{28.97}{1.89\times 10^{-4}} &	\withrelerr{18.06}{1.62\times 10^{-3}} &	 \withrelerr{14.99}{1.31\times 10^{-2}} &	\withrelerr{14.46}{4.24\times 10^{-2}} \\
\hline
\end{tabular}
\caption{\emph{Computing time in seconds as a function of the approximation level and the grid size. The numbers below indicate the relative error on the non-vanilla put price with respect to the best estimate.}}
\label{conv-nv-time}
\end{table}

\begin{table}[p]
\centering
\begin{tabular}{|c|rrrrr|}
\hline
 & \multicolumn{1}{c}{0\,\%}&\multicolumn{1}{c}{5\,\%} & \multicolumn{1}{c}{20\,\%} & \multicolumn{1}{c}{50\,\%} & \multicolumn{1}{c|}{100\,\%} \\
\hline
50   & $1.35\times 10^{+1}$	 & $1.35\times 10^{+1}$	 & $1.36\times 10^{+1}$	 & $1.37\times 10^{+1}$	 & $1.40\times 10^{+1}$ \\
100  & $-7.36\times 10^{-2}$ & $-7.34\times 10^{-2}$ & $-7.20\times 10^{-2}$ & $-6.07\times 10^{-2}$ & $-3.22\times 10^{-2}$ \\
250  & $-4.69\times 10^{-2}$ & $-4.67\times 10^{-2}$ & $-4.53\times 10^{-2}$ & $-3.43\times 10^{-2}$ & $-5.94\times 10^{-3}$ \\
500  & $-1.21\times 10^{-2}$ & $-1.19\times 10^{-2}$ & $-1.05\times 10^{-2}$ & $7.52\times 10^{-4}$	 & $3.12\times 10^{-2}$ \\
750  & $-4.96\times 10^{-3}$ & $-4.77\times 10^{-3}$ & $-3.30\times 10^{-3}$ & $8.45\times 10^{-3}$	 & $3.83\times 10^{-2}$ \\
1000 & $-2.53\times 10^{-3}$ & $-2.34\times 10^{-3}$ & $-8.90\times 10^{-4}$ & $1.07\times 10^{-2}$	 & $4.00\times 10^{-2}$ \\
1500 & $-6.99\times 10^{-4}$ & $-5.11\times 10^{-4}$ & $9.28\times 10^{-4}$	 & $1.24\times 10^{-2}$	 & $4.15\times 10^{-2}$ \\
2000 & $-1.75\times 10^{-4}$ & $1.47\times 10^{-5}$	 & $1.46\times 10^{-3}$	 & $1.29\times 10^{-2}$	 & $4.22\times 10^{-2}$ \\
2500 & $$                    & $1.89\times 10^{-4}$  & $1.62\times 10^{-3}$	 & $1.31\times 10^{-2}$  & $4.24\times 10^{-2}$ \\
\hline
\end{tabular}
\caption{\emph{Relative errors for the price of the put on the non-vanilla CPPI as a function of the approximation level and the grid size.}}
\label{conv-nv-price}
\end{table}

\begin{table}[p]
\centering
\begin{tabular}{|c|rrrrr|}
\hline
 & \multicolumn{1}{c}{0\,\%}&\multicolumn{1}{c}{5\,\%} & \multicolumn{1}{c}{20\,\%} & \multicolumn{1}{c}{50\,\%} & \multicolumn{1}{c|}{100\,\%} \\
\hline
50   & $1.55\times 10^{+1}$  & $1.55\times 10^{+1}$  & $1.56\times 10^{+1}$	& $1.56\times 10^{+1}$	& $1.62\times 10^{+1}$ \\
100	 & $-6.99\times 10^{-2}$ & $-6.98\times 10^{-2}$ & $-6.91\times 10^{-2}$	& $-6.26\times 10^{-2}$	& $-5.27\times 10^{-2}$ \\
250	 & $-5.10\times 10^{-2}$ & $-5.09\times 10^{-2}$ & $-5.04\times 10^{-2}$	& $-4.51\times 10^{-2}$	& $-3.47\times 10^{-2}$ \\
500	 & $-1.44\times 10^{-2}$ & $-1.43\times 10^{-2}$ & $-1.38\times 10^{-2}$	& $-8.53\times 10^{-3}$	& $1.30\times 10^{-3}$ \\
750	 & $-5.54\times 10^{-3}$ & $-5.48\times 10^{-3}$ & $-4.95\times 10^{-3}$	& $1.40\times 10^{-4}$	& $1.03\times 10^{-2}$ \\
1000 & $-2.76\times 10^{-3}$ & $-2.69\times 10^{-3}$ & $-2.16\times 10^{-3}$	& $3.01\times 10^{-3}$	& $1.34\times 10^{-2}$ \\
1500 & $-9.01\times 10^{-4}$ & $-8.33\times 10^{-4}$ & $-2.94\times 10^{-4}$	& $4.92\times 10^{-3}$	& $1.54\times 10^{-2}$ \\
2000 & $-2.85\times 10^{-4}$ & $-2.17\times 10^{-4}$ & $3.21\times 10^{-4}$	& $5.53\times 10^{-3}$	& $1.60\times 10^{-2}$ \\
2500 & $$                    & $6.82\times 10^{-5}$	& $6.12\times 10^{-4}$	& $5.83\times 10^{-3}$	& $1.63\times 10^{-2}$ \\
\hline
\end{tabular}
\caption{\emph{Relative errors for the delta of the put on the non-vanilla CPPI as a function of the approximation level and the grid size.}}
\label{conv-nv-delta}
\end{table}

\FloatBarrier

\begin{table}[htbp]
\centering
\begin{tabular}{|c|rrrrr|}
\hline
 & \multicolumn{1}{c}{0\,\%}&\multicolumn{1}{c}{5\,\%} & \multicolumn{1}{c}{20\,\%} & \multicolumn{1}{c}{50\,\%} & \multicolumn{1}{c|}{100\,\%} \\
\hline
50	 &  $-1.84\times 10^{+0}$ &	$-1.84\times 10^{+0}$ &	$-1.79\times 10^{+0}$ &	$-1.91\times 10^{+0}$ &	$-2.32\times 10^{+0}$ \\
100	 &  $-7.25\times 10^{-2}$ &	$-7.23\times 10^{-2}$ &	$-7.11\times 10^{-2}$ &	$-6.12\times 10^{-2}$ &	$-3.35\times 10^{-2}$ \\
250	 &  $-4.67\times 10^{-2}$ &	$-4.66\times 10^{-2}$ &	$-4.53\times 10^{-2}$ &	$-3.59\times 10^{-2}$ &	$-1.00\times 10^{-2}$ \\
500	 &  $-1.28\times 10^{-2}$ &	$-1.27\times 10^{-2}$ &	$-1.14\times 10^{-2}$ &	$-1.75\times 10^{-3}$ &	$2.58\times 10^{-2}$ \\
750  &  $-5.22\times 10^{-3}$ &	$-5.05\times 10^{-3}$ &	$-3.76\times 10^{-3}$ &	$6.33\times 10^{-3}$  &	$3.35\times 10^{-2}$ \\
1000 &	$-2.65\times 10^{-3}$ &	$-2.48\times 10^{-3}$ &	$-1.20\times 10^{-3}$ &	$8.74\times 10^{-3}$  &	$3.55\times 10^{-2}$ \\
1500 &	$-7.32\times 10^{-4}$ &	$-5.66\times 10^{-4}$ &	$7.03\times 10^{-4}$  & $1.06\times 10^{-2}$  &	$3.72\times 10^{-2}$ \\
2000 &	$-1.82\times 10^{-4}$ &	$-1.44\times 10^{-5}$ &	$1.26\times 10^{-3}$  &	$1.11\times 10^{-2}$  &	$3.79\times 10^{-2}$ \\
2500 &	$$                    & $1.66\times 10^{-4}$  &	$1.43\times 10^{-3}$ &	$1.13\times 10^{-2}$ & $3.80\times 10^{-2}$ \\
\hline
\end{tabular}
\caption{\emph{Relative errors for the vega of the put on the non-vanilla CPPI as a function of the approximation level and the grid size.}}
\label{conv-nv-vega}
\end{table}

\begin{table}[htbp]
\centering
\begin{tabular}{|c|rrrrr|}
\hline
 & \multicolumn{1}{c}{0\,\%}&\multicolumn{1}{c}{5\,\%} & \multicolumn{1}{c}{20\,\%} & \multicolumn{1}{c}{50\,\%} & \multicolumn{1}{c|}{100\,\%} \\
\hline
50	 &  $-1.29\times 10^{+1}$ &	$-1.29\times 10^{+1}$ &	$-1.29\times 10^{+1}$ &	$-1.25\times 10^{+1}$ &	$-9.52\times 10^{+0}$ \\
100	 &  $-6.85\times 10^{-2}$ &	$-6.48\times 10^{-2}$ &	$-4.92\times 10^{-2}$ &	$7.58\times 10^{-3}$ &	$7.34\times 10^{-2}$ \\
250	 &  $4.11\times 10^{-2}$ &	$4.27\times 10^{-2}$ &	$5.47\times 10^{-2}$ &	$1.14\times 10^{-1}$ &	$3.44\times 10^{-1}$ \\
500	 &  $4.83\times 10^{-2}$ &	$5.05\times 10^{-2}$ &	$6.35\times 10^{-2}$ &	$1.28\times 10^{-1}$ &	$4.43\times 10^{-1}$ \\
750  &  $1.36\times 10^{-2}$ &	$1.50\times 10^{-2}$ &	$2.56\times 10^{-2}$ &	$1.07\times 10^{-1}$  &	$3.72\times 10^{-1}$ \\
1000 &	$5.37\times 10^{-3}$ &	$6.02\times 10^{-3}$ &	$1.67\times 10^{-2}$ &	$8.96\times 10^{-2}$  &	$3.21\times 10^{-1}$ \\
1500 &	$2.04\times 10^{-3}$ &	$3.67\times 10^{-3}$ &	$1.28\times 10^{-2}$  & $8.19\times 10^{-2}$  &	$3.01\times 10^{-1}$ \\
2000 &	$1.67\times 10^{-4}$ &	$1.12\times 10^{-3}$ &	$1.24\times 10^{-2}$  &	$8.17\times 10^{-2}$  &	$3.08\times 10^{-1}$ \\
2500 &	$$                   &  $2.41\times 10^{-3}$ &	$9.33\times 10^{-3}$ &	$7.84\times 10^{-2}$ & $3.06\times 10^{-2}$ \\
\hline
\end{tabular}
\caption{\emph{Relative errors for the gap proportion of the non-vanilla CPPI as a function of the approximation level and the grid size.}}
\label{conv-nv-gp}
\end{table}

\begin{figure}[htbp]
\centering
\includegraphics[width=.75\textwidth]{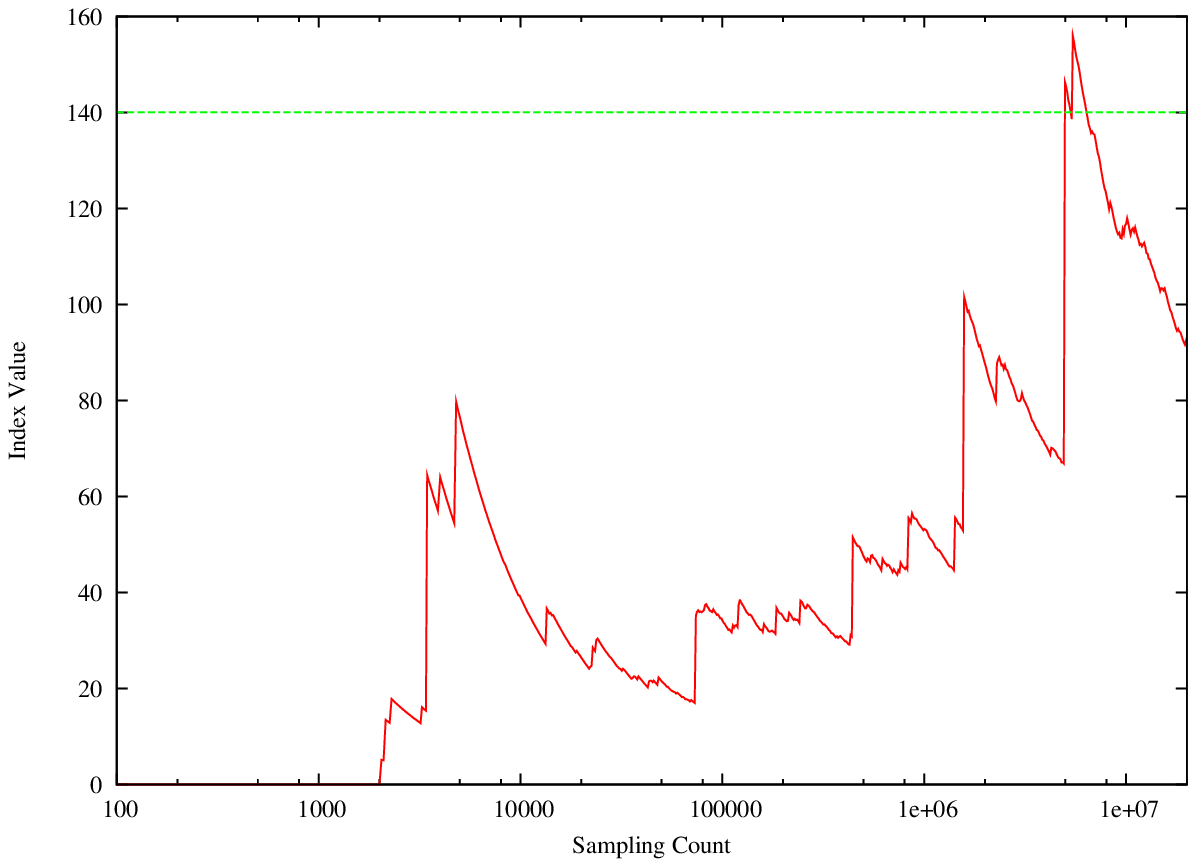}
\includegraphics[width=.75\textwidth]{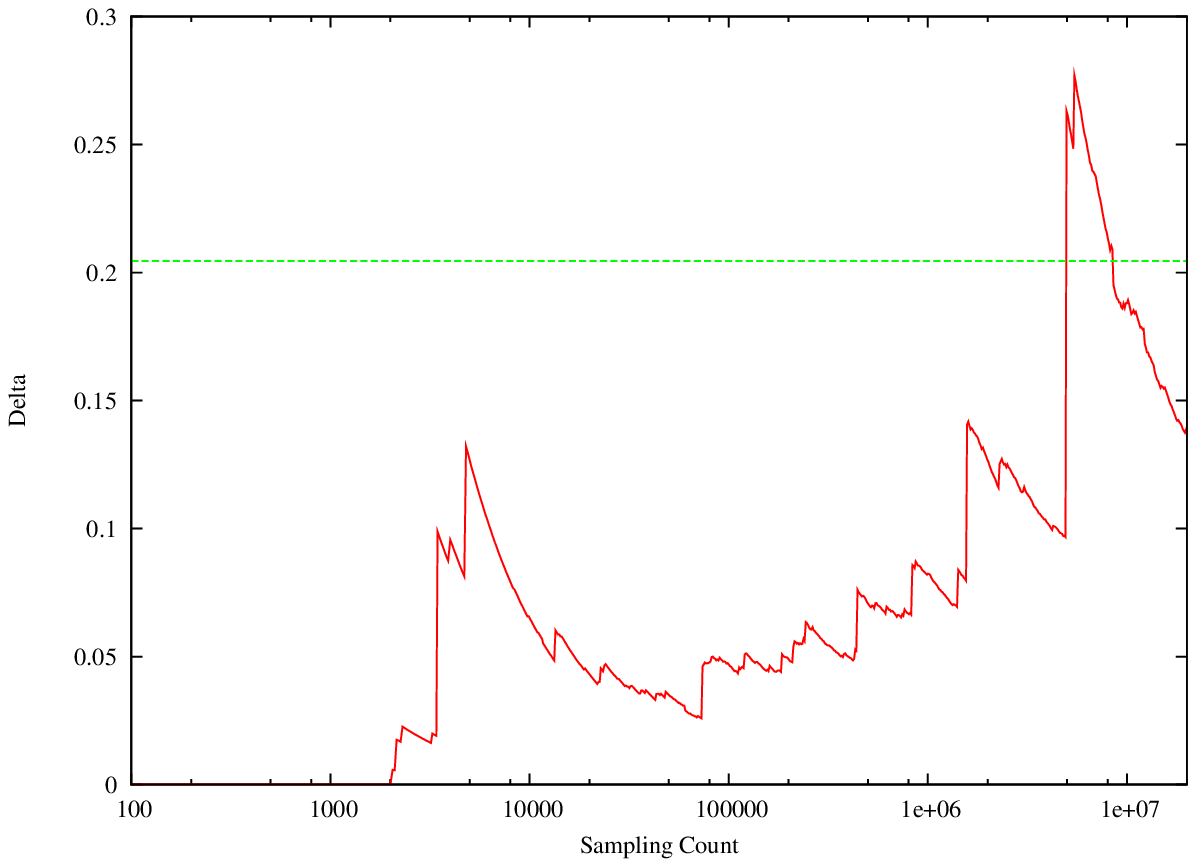}
\includegraphics[width=.75\textwidth]{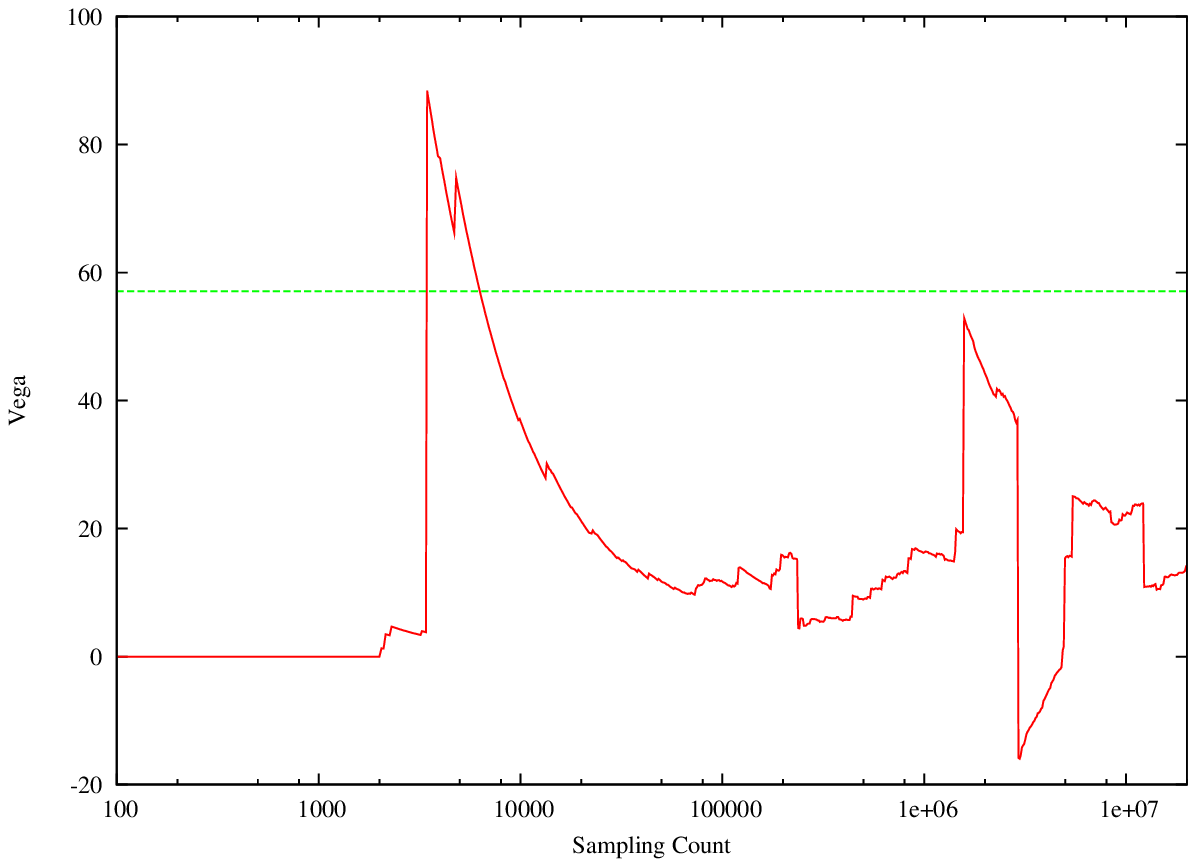}
\caption{\emph{Convergence of the Monte Carlo method for price, delta and vega in the non-vanilla case, as functions of the number of paths. In each graph, the dashed line indicates the most accurate value obtained with the Markov scheme.}}
\label{fig-nonvanilla-mc}
\end{figure}

For comparison, the convergence graphs of the Monte-Carlo pricing are shown on figure \ref{fig-nonvanilla-mc}. As in the vanilla case, the price and greeks have not converged after 20 000 000 paths and 18 hours of computation.

\subsection{Profit lock-in}

Computing times are higher when profit lock-in is used. In this case, matrix-matrix multiplications have to be performed instead of matrix-vector multiplications. In addition, transition matrices over lock-in period have to be transformed, which can take some time. Both aspects make the algorithm $O(N^3)$ instead of $O(N^2)$ where $N$ is the grid size. For example, the $N=500$ case with 20\% tolerance on the rate variation takes 26.7\,s instead of 510\,ms. However this is still reasonable and much better than a Monte-Carlo pricing. Parallelization of the algorithm would also reduce the computation time.

\FloatBarrier

\section{Features analysis}
\label{section-features-model}

In order to analyze various features of CPPI strategies, we consider now a 10 years CPPI with weekly rebalancing on some risky asset with 20\% volatility. The multiplier is 4 and there are no fees nor spreads on the risk-free asset. The risk-free interest is taken constant at 5\% in continuous compounding. In a perspective of pricing and hedging the gap risk, we consider the risk-neutral probability. Other probability spaces could be used for other purposes.

For each case, we compute the following gap indicators
\begin{itemize}
\item {\bf Gap Proportion:} probability of the CPPI strategy being below the guarantee at maturity, $\mathbb{P} \left[ C_T < 1 \right]$;
\item {\bf Conditional Loss:} mean value of the loss when the strategy ends below the guarantee, $\mathbb{E}[(1-C_T)^+ \, \vert \, C_T < 1]$;
\item {\bf Expected Loss:} expected value of the loss without conditioning, $\mathbb{E}[(1-C_T)^+]$.
\end{itemize}

\subsection{Risky asset model}

The model of the risky asset has a large influence on the gap risk. We have performed computations for three cases:
\begin{itemize}
\item Brownian motion with 20\% constant volatility;
\item Jump-diffusion, adding exponentially decreasing jumps to the Brownian motion in both directions (Kou model). The volatility of the Brownian is still $\sigma = 20\%$ and the jumps parameter set is $\lambda_+ = 0.1$, $\eta_+ = 0.05$, $\lambda_- = 0.1$,$\eta_- = 0.1$. The total volatility of the process is $\sqrt{\sigma^2 + 2\lambda_+ \eta_+^2 + 2\lambda_- \eta_-^2} \simeq 21.6795\%$.
\item Brownian motion which gives the same expected loss $\mathbb{E}[(1-C_T)^+]$ as the jump diffusion case. The volatility which fits this number is close to 62\%.
\end{itemize}

\begin{figure}[htbp]
\centering
\includegraphics[width=\textwidth]{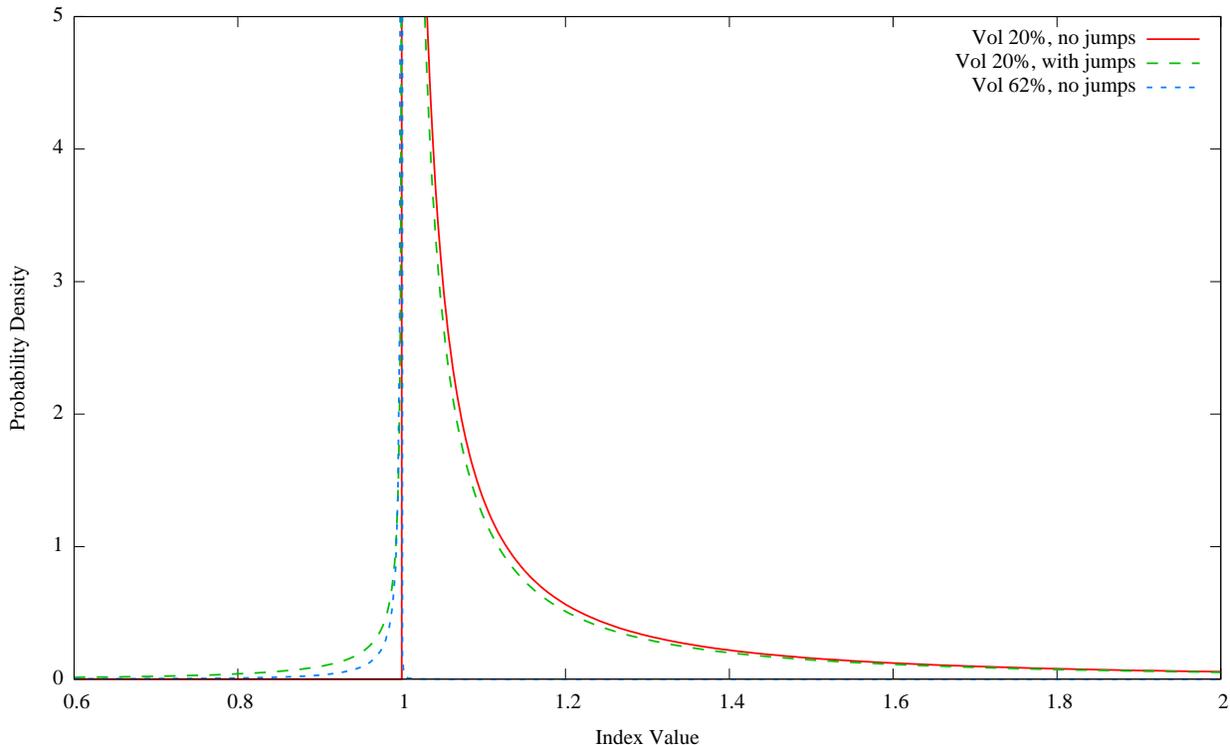}
\caption{\emph{Probability Density of a CPPI strategy final value depending on the risky asset model. CPPI has maturity 10 years, weekly rebalancing, multiplier 4. The initial investment and the guaranteed level are 1. The risk-free interest rate is 5\%, so the mean price of the strategy is 1.65. There are no fees. The first curve is the probability density for a risky asset following a Brownian motion with 20\% constant volatility. for the second curves, jumps have been added: downward jump with intensity 0.1 and mean value 0.1, upward jumps with intensity 0.1 and mean value 0.05. The third curve corresponds to a Brownian motion without jumps which gives the same gap risk, \emph{i.e.} with 62\% volatility.}}
\label{fig-risky}
\end{figure}

As shown in table \ref{tab-risky} and figure \ref{fig-risky}, adding jumps changes completely the loss. With a 20\% volatility in pure Brownian motion, there is no loss probability up to numerical accuracy ($\sim~10^{-14}$). With multiplier 4, a loss occurs when the risky asset price drops by more than 25\%, which have negligible probability in one week at the volatility considered. Adding jumps modifies completely this behavior and results in a completely different situation, more consistent with observed skews. Extreme moves are allowed with some small probability. With the parameters that we have taken, this gives a 5\% probability of breaking the threshold in ten years, which is not negligible.

\begin{table}[htbp]
\centering
\begin{tabular}{|l|r|r|r|}
\hline
Model & Gap Prop. & Cond. Loss & Expect. Loss \\
\hline
Brownian 20\,\% &  0.00\,\% &  0.00\,\% & 0.000\,\% \\
Kou 20\,\%      &  5.71\,\% & 18.41\,\% & 1.052\,\% \\
Brownian 62\,\% & 21.78\,\% &  4.85\,\% & 1.058\,\% \\
\hline
\end{tabular}
\caption{\emph{Gap proportion, conditional loss and expected loss depending on the risky asset model.}}
\label{tab-risky}
\end{table}

In an attempt to take into account the gap risk using pure Brownian motion, one could be tempted to increase the model volatility. We find that the volatility which would give the same expected loss would be as high as 62\%. However it results in a completely different probability distribution for the strategy price. The loss are much more probable but have a much lower mean value: the left tails have different shapes. The right part of the graph depends mainly on the variance of the underlying process and is almost given by a lognormal law, shifted by 1, centered on 1.65 (the forward value with 5\% interest rate) and with standard deviation $m\sigma\sqrt{T}$ where $m$ is the multiplier and $\sigma\sqrt{T}$ the square root of the variance of the underlying. Therefore the probability density with 62\% volatility has nothing to do with the 20\% case neither in the gap part of the distribution nor in the well-performing part. Increasing the model volatility is not a solution for pricing or hedging the gap risk.

For equities, jump-diffusion is more realistic model than pure Brownian motion: it is more consistent a non-vanishing short-term skew and historical returns seems to be consistent with Kou jump-diffusion model \cite{cont:cpp}. Therefore we consider in the following only this case, with the volatility and jump parameters given above.

\FloatBarrier
\subsection{Multiplier}

\begin{figure}[p]
\centering
\includegraphics[width=\textwidth]{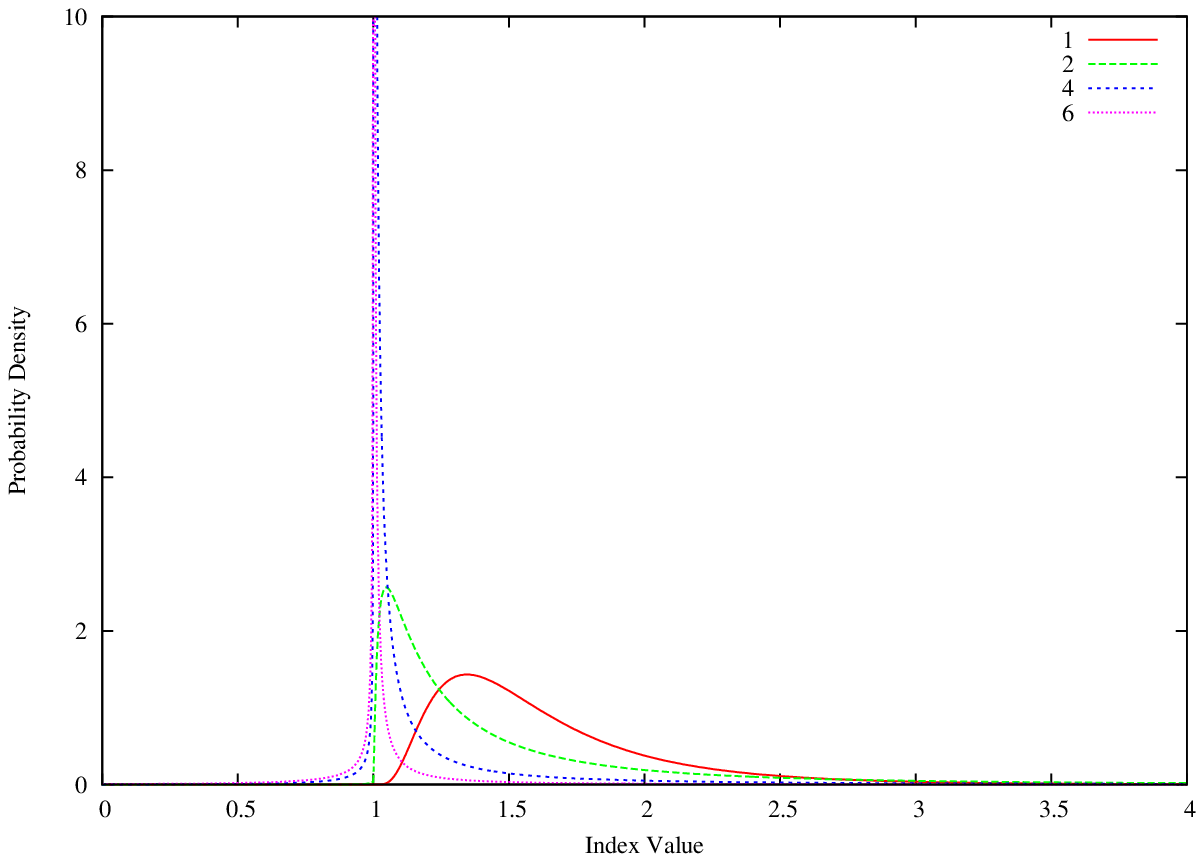}
\includegraphics[width=\textwidth]{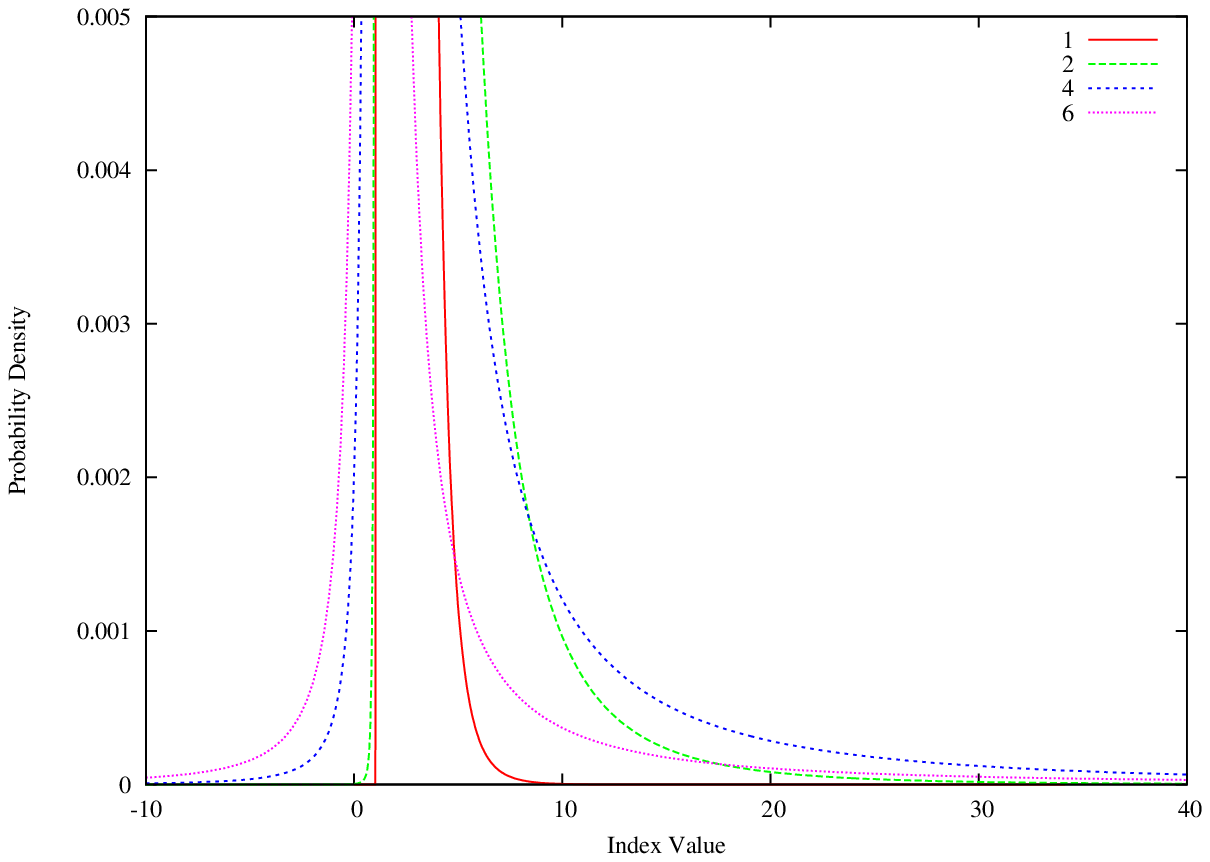}
\caption{\emph{Probability Density of the CPPI strategy depending on the multiplier. The underlying is the Kou model which includes jumps with the same parameters as in figure \ref{fig-risky}. The first graph shows the shape of curves near the threshold. The second graph shows the tails of the distributions.}}
\label{fig-multiplier}
\end{figure}

The multiplier increases the variance of the CPPI strategy by leveraging the risky asset. As a consequence a big multiplier increases the gap risk and the probability of very high returns, as it can be seen on table \ref{tab-multiplier} and figure \ref{fig-multiplier}.
\begin{table}[htbp]
\centering
\begin{tabular}{|l|r|r|r|}
\hline
Multiplier & Gap Prop. & Cond. Loss & Expect. Loss  \\
\hline
1 &  0.00\,\% &  0.00\,\% & 0.000\,\% \\
2 &  0.10\,\% &  5.91\,\% & 0.605\,\% \\
4 &  5.71\,\% & 18.41\,\% & 1.052\,\% \\
6 & 15.51\,\% & 33.37\,\% & 5.177\,\% \\
\hline
\end{tabular}
\caption{\emph{Gap proportion, conditional loss and expected loss depending on the multiplier.}}
\label{tab-multiplier}
\end{table}
The case of unit multiplier is degenerate: there is no leverage and no need to rebalance. The strategy reduces to a zero-coupon plus some shares which are kept until the maturity. It can be checked that in this case there is no gap risk: the probability density is null below 1.

The floor is broken when there is a downward move larger than the inverse of the multiplier between two rebalancing dates: $\frac{\Delta S}{S}<-\frac{1}{m}$. For a multiplier larger than 1, there is a positive probability of losing more than the initial investment. Because of the leverage, the CPPI strategy can end with a negative value when a downward jump in the risky asset is larger than the financing component.


\subsection{Maximum exposure}

In order to reduce the exposition to adverse moves and reduce the gap risk, an efficient way is setting a maximum exposure, which is a cap on the risky asset weighting. When there is no such constraint, the asymptotic value of the exposure is given by the multiplier. For the same CPPI as above, we show in table \ref{tab-expomax} the effect on the gap risk of reducing the maximum exposure from the unconstrained 400\% case (recall that $m=4$ in our example).

\begin{table}[htbp]
\centering
\begin{tabular}{|l|r|r|r|}
\hline
Max. Expo. & Gap Prop. & Cond. Loss & Expect. Loss \\
\hline
400\,\% & 5.71\,\% & 18.41\,\% & 1.052\,\% \\
200\,\% & 5.16\,\% &  8.13\,\% & 0.419\,\% \\
150\,\% & 4.44\,\% &  5.88\,\% & 0.261\,\% \\
100\,\% & 2.92\,\% &  3.66\,\% & 0.107\,\% \\
\hline
\end{tabular}
\caption{\emph{Gap proportion, conditional loss and expected loss depending on the maximum exposure.}}
\label{tab-expomax}
\end{table}

\begin{figure}[p]
\centering
\includegraphics[width=\textwidth]{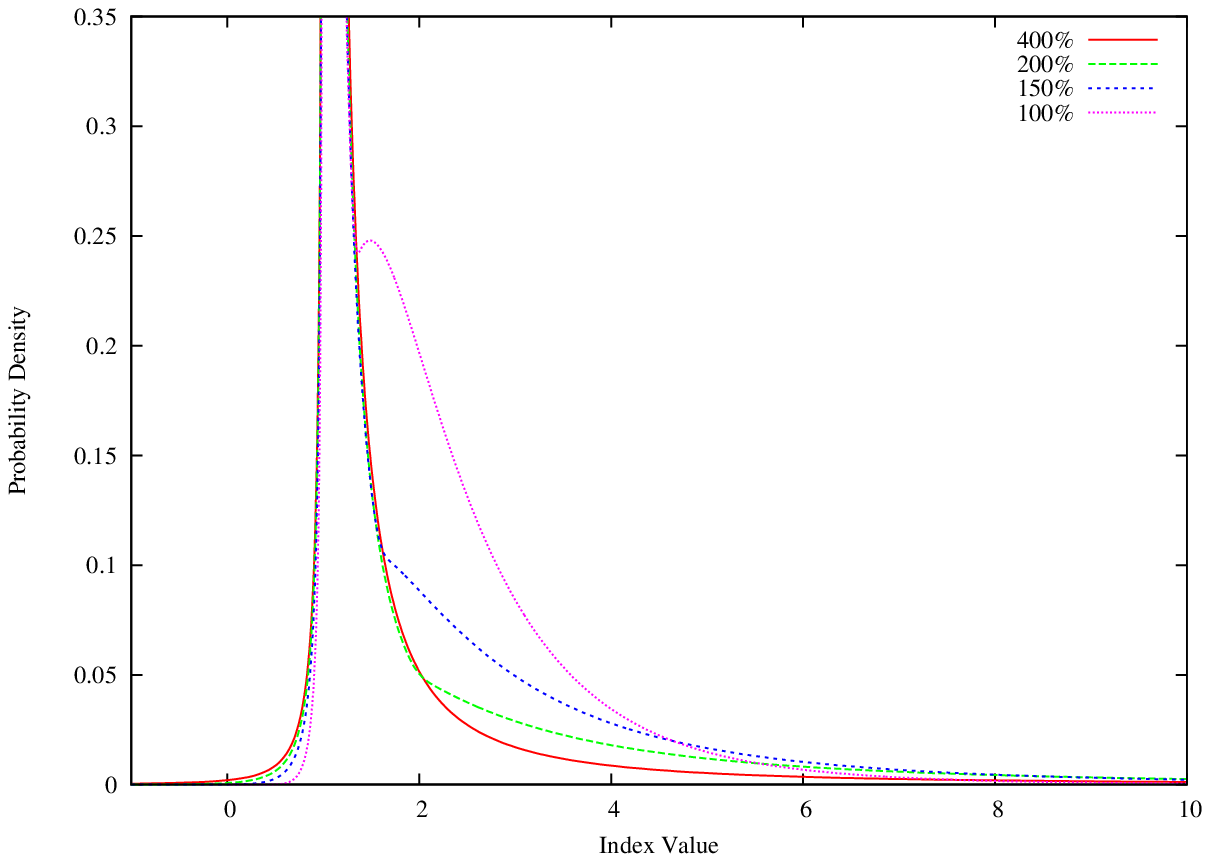}
\includegraphics[width=\textwidth]{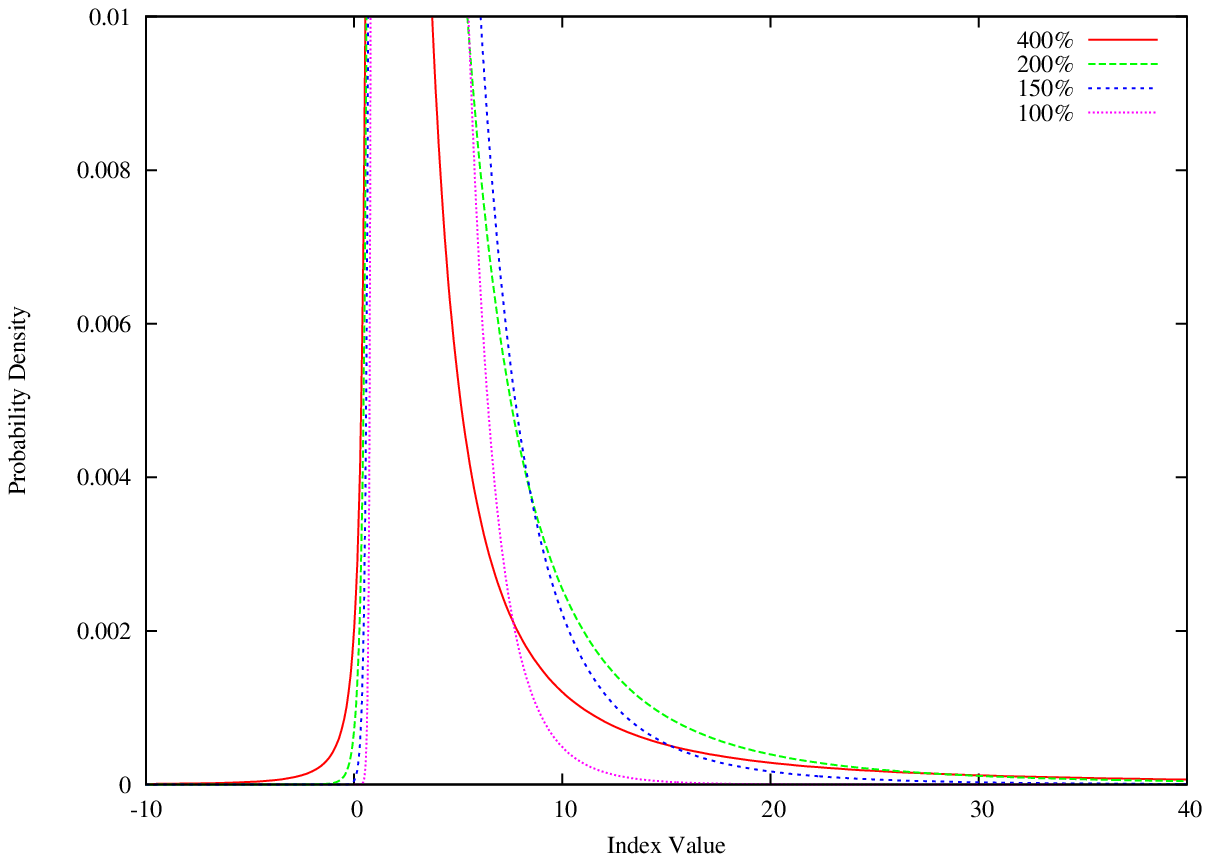}
\caption{\emph{Probability Density of the CPPI strategy depending on the maximum exposure. The multiplier is 4. The first graph shows the shape of curves near the threshold. The second graph shows the tails of the distributions.}}
\label{fig-expomax}
\end{figure}

When a loss occurs on the CPPI strategy, its value is proportional to the index level and the multiplier. Therefore bounding the exposure decreases the probability of a large loss by capping the exposure when the index level is high. The probability of little losses are less modified: the left tail is shortened. The effects on both tails of the distribution can be seen in figure \ref{fig-expomax}. There is a slope change at the level where the maximum exposure is reached. The power effect of the CPPI strategy is stopped and the right part of the distribution is more or less a dilated version of the underlying probability density. The second graph shows that both tails are reduced; in particular limiting the exposure to 100\% obviously prevent the CPPI strategy to drop below 0.

\FloatBarrier

\subsection{Minimum exposure}

\begin{figure}[htbp]
\centering
\includegraphics[width=\textwidth]{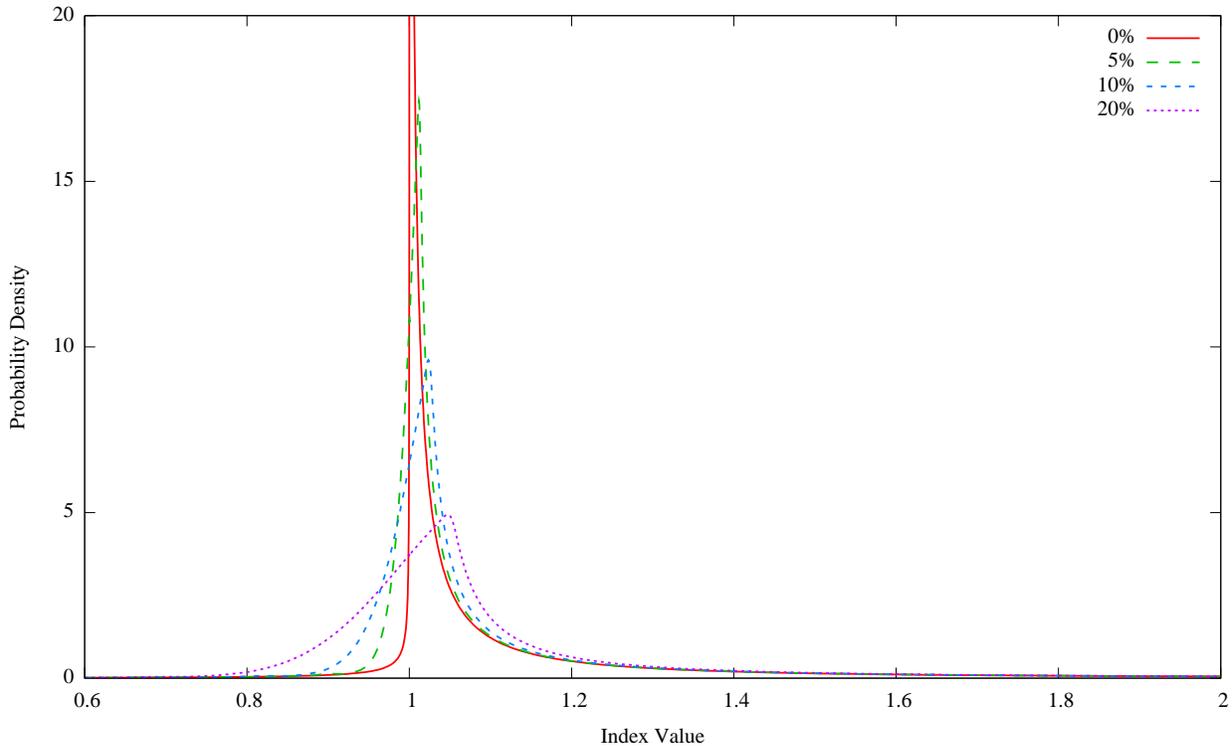}
\caption{\emph{Probability Density of the CPPI strategy depending on the minimum exposure. The tails are not shown as they are almost identical.}}
\label{fig-expomin}
\end{figure}

One may want to set a minimum exposure, \emph{i.e.} a floor on the risky asset weighting, so that there is always some possibility to recover from a gap event. Figure \ref{fig-expomin} shows that this features does not change the CPPI distribution above some value which is not very high, here 1.2, to be compared with the mean value which is 1.65.
The gap risk increases when the minimum exposure gets higher, as shown in table \ref{tab-expomin}. Some small diffusion losses appear, which results in a large gap proportion but a relatively low increase of the expected loss. Large losses come from jumps starting at high levels, where the distribution is not changed.

\begin{table}[htbp]
\centering
\begin{tabular}{|l|r|r|r|}
\hline
Min. Expo. & Gap Prop. & Cond. Loss & Expect. Loss \\
\hline
0\,\%  &  5.71\,\% & 18.41\,\% & 1.052\,\% \\
5\,\%  & 19.77\,\% &  6.39\,\% & 1.263\,\% \\
10\,\% & 24.87\,\% &  6.70\,\% & 1.666\,\% \\
20\,\% & 30.80\,\% &  8.96\,\% & 2.759\,\% \\
\hline
\end{tabular}
\caption{\emph{Gap proportion, conditional loss and expected loss depending on the minimum exposure.}}
\label{tab-expomin}
\end{table}

\FloatBarrier

\subsection{Cushion limit}
\label{section-cushion-limit}

Capping the exposure to some maximum value reduces the gap risk. An other feature which can prevent some losses is the cushion limit: the exposure is set to zero below some cushion value. Gap indicators of table \ref{tab-cushionlimit} show that this effectively reduces the gap proportion: the cases where the CPPI strategy fall below the limit value are frozen and cannot defease any longer. However the expected loss is not significantly reduced.

\begin{table}[htbp]
\centering
\begin{tabular}{|l|r|r|r|}
\hline
Limit & Gap Prop. & Cond. Loss & Expect. Loss \\
\hline
0\,\%  & 5.71\,\% & 18.41\,\% & 1.052\,\% \\
3\,\%  & 3.95\,\% & 26.22\,\% & 1.036\,\% \\
8\,\%  & 2.92\,\% & 33.83\,\% & 0.988\,\% \\
15\,\% & 2.05\,\% & 43.05\,\% & 0.884\,\% \\
\hline
\end{tabular}
\caption{\emph{Gap proportion, conditional loss and expected loss depending on the cushion limit.}}
\label{tab-cushionlimit}
\end{table}

\begin{figure}[htbp]
\centering
\includegraphics[width=\textwidth]{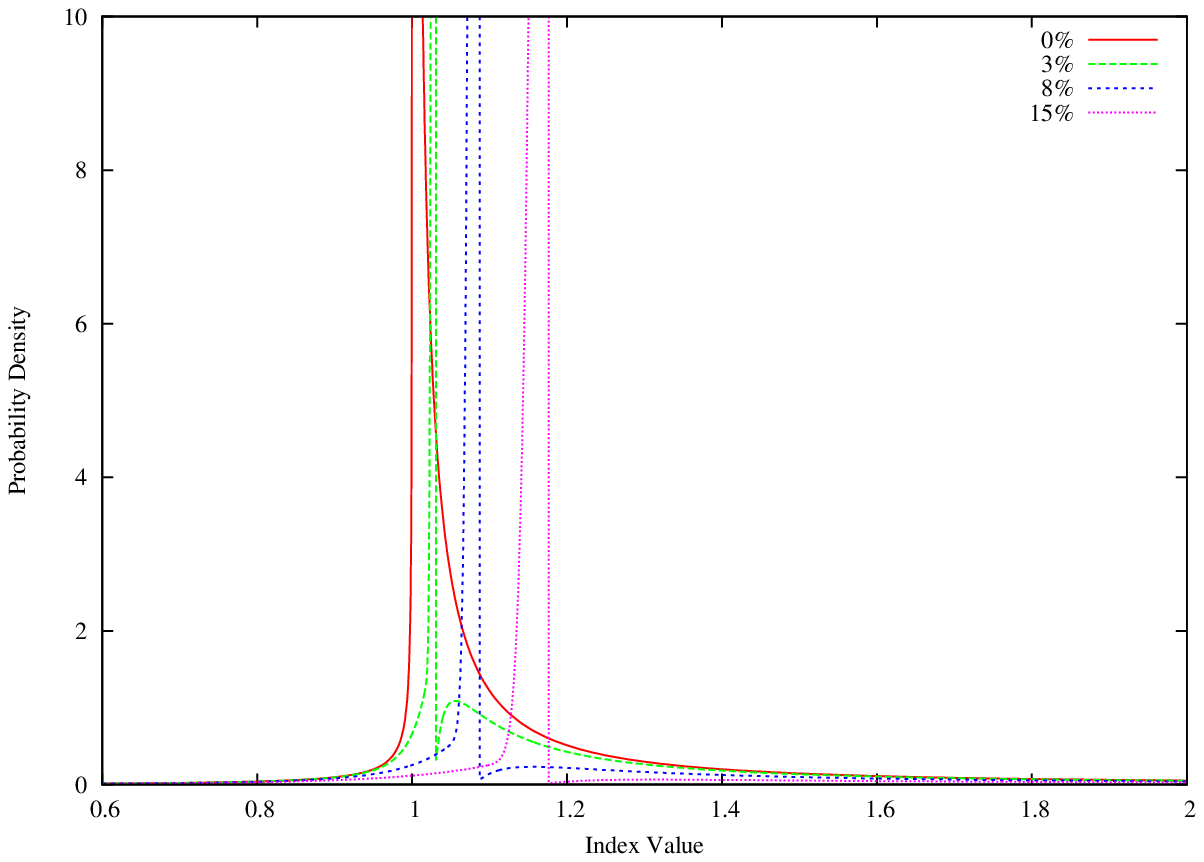}
\includegraphics[width=\textwidth]{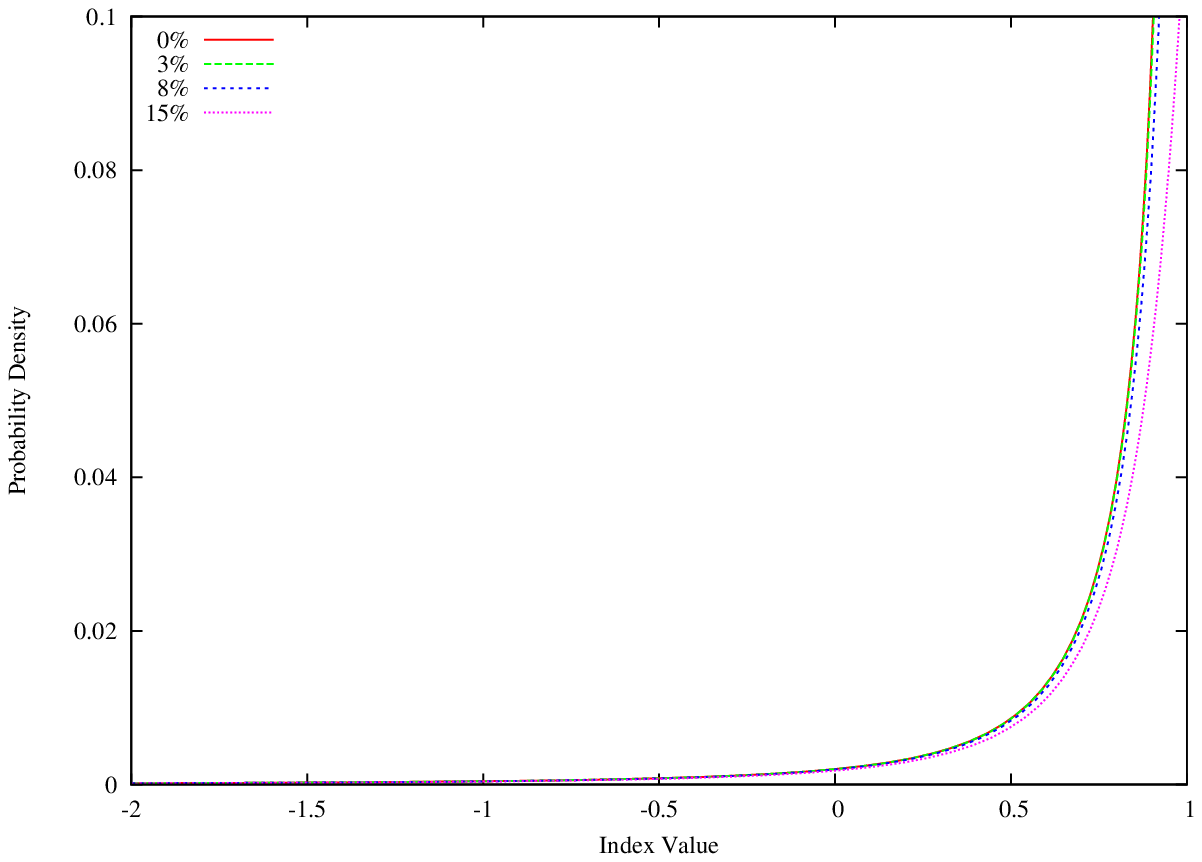}
\caption{\emph{Probability Density of the CPPI strategy depending on the cushion limit. The second graph shows the left tail of the distributions. The curves are almost identical on the right tail.}}
\label{fig-cushionlimit}
\end{figure}

The effect of this feature on the distribution can be observed on figure \ref{fig-cushionlimit}: the Dirac-like probability peak just above the threshold is displaced because the motion is stopped before low values are reached. Losses due to large jumps from higher values are not modified.

\subsection{Profit lock-in}

The profit lock-in feature locks some performance at periodic intervals. We consider here annual lock-in and look at the effect of the lock-in proportion. Table \ref{tab-lockin} presents two series of indicators. In the first set of column, the guarantee is kept at the initial level whereas the threshold value used to compute the cushion is increased. The gap risk is therefore reduced because the lock-in only reduces the exposure. The second set of column deals with the case where the guaranteed level is increased of the lock-in amount. The gap proportion is not changed in this case, as the CPPI threshold and guarantee are increased in the same proportion. However the expected loss is reduced because the exposure is reduced.

\begin{table}[htbp]
\centering
\begin{tabular}{|l||r|r|r||r|r|r|}
\hline
& \multicolumn{3}{c||}{Initial Guarantee} & \multicolumn{3}{c|}{Locked-In Guarantee} \\
\hline
Prop. & Gap Prop. & Cond. Loss & Exp. Loss & Gap Prop. & Cond. Loss & Exp. Loss \\
\hline
0\,\%  & 5.71\,\% & 18.41\,\% & 1.052\,\% & 5.71\,\% & 18.41\,\% & 1.052\,\% \\
25\,\% & 2.20\,\% & 18.95\,\% & 0.417\,\% & 5.71\,\% & 13.61\,\% & 0.777\,\% \\
50\,\% & 1.72\,\% & 14.57\,\% & 0.251\,\% & 5.71\,\% & 10.21\,\% & 0.583\,\% \\
75\,\% & 1.52\,\% & 12.91\,\% & 0.197\,\% & 5.71\,\% &  7.84\,\% & 0.448\,\% \\
\hline
\end{tabular}
\caption{\emph{Gap proportion, conditional loss and expected loss depending on the Lock-In proportion, with respect to the initial guarantee and with respect to the final locked-in guarantee.}}
\label{tab-lockin}
\end{table}

\begin{figure}[htb]
\centering
\includegraphics[width=\textwidth]{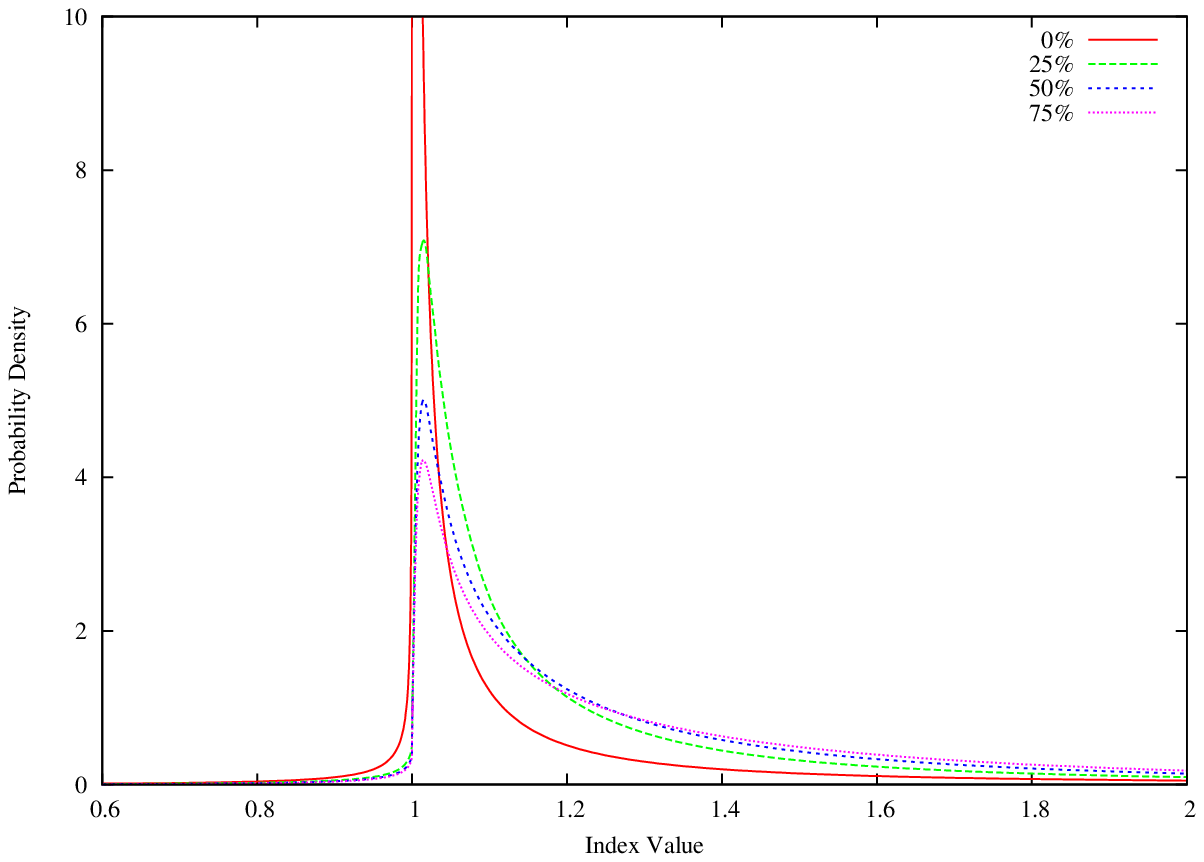}
\includegraphics[width=\textwidth]{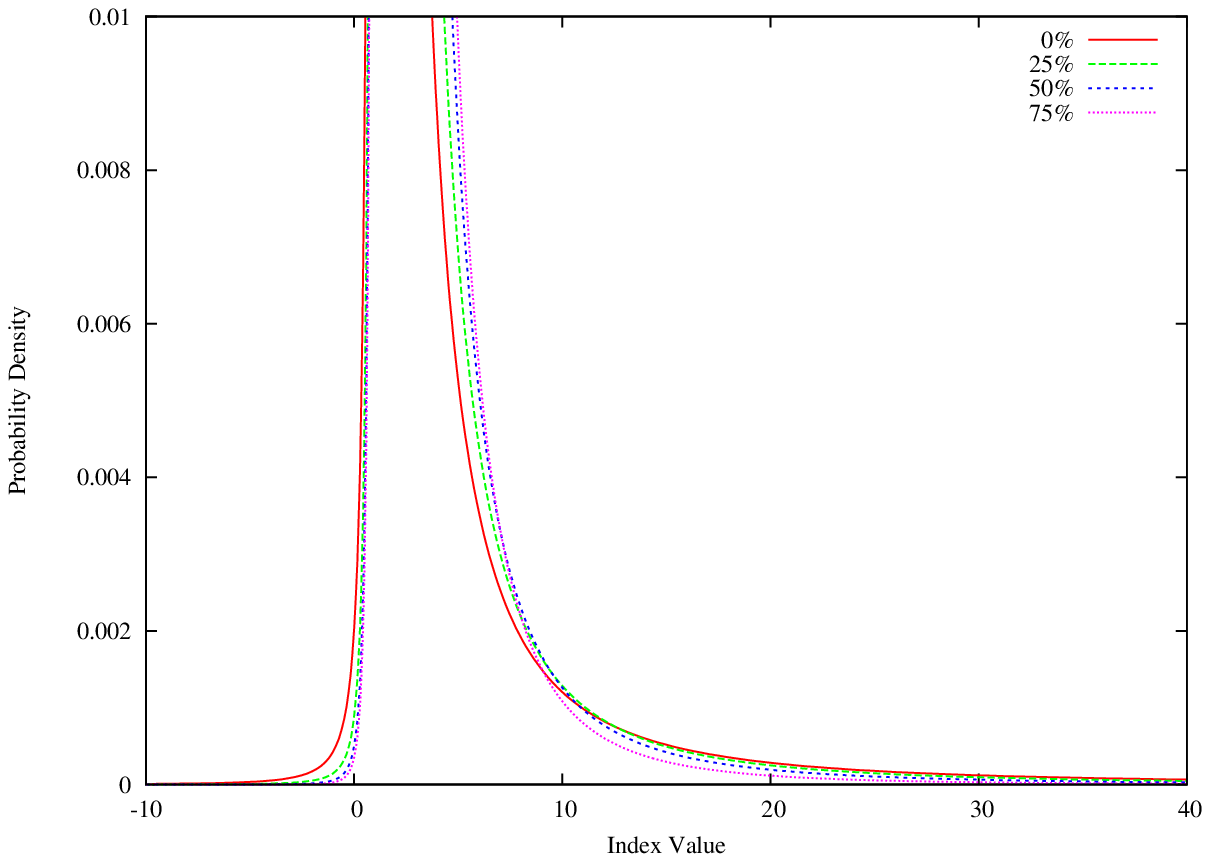}
\caption{\emph{Probability Density of the CPPI strategy depending on the profit lock-in proportion. Annual performance is locked with the given proportion. The second graph shows the tails of the distributions.}}
\label{fig-lockin}
\end{figure}

Figure \ref{fig-lockin} shows the effect on the curve and on its tails which are reduced by lowering the leverage. The effect is smoother than with a maximum exposure, but if the lock-in applied to the guarantee it does not reduce the gap risk as efficiently.

\subsection{Artificial threshold}

Finally we discuss the effect of artificial thresholds: the threshold is no longer defined by the natural zero-coupon value. Either it is defined by a linear function of time, by a fixed rate, or by a spread over the zero-coupon curve. We focus here on the latter case. With a negative spread, the initial threshold is higher than the natural one and finishes at the the level. The exposure is thus reduced, especially for low values of the CPPI index. This moves slightly a part of the distribution which was close to the threshold and smoothes the peak of the distribution. On the other hand, the right tail is also reduced as the mean value remain the same, here 1.65.

\FloatBarrier 

\begin{figure}[htbp]
\centering
\includegraphics[width=\textwidth]{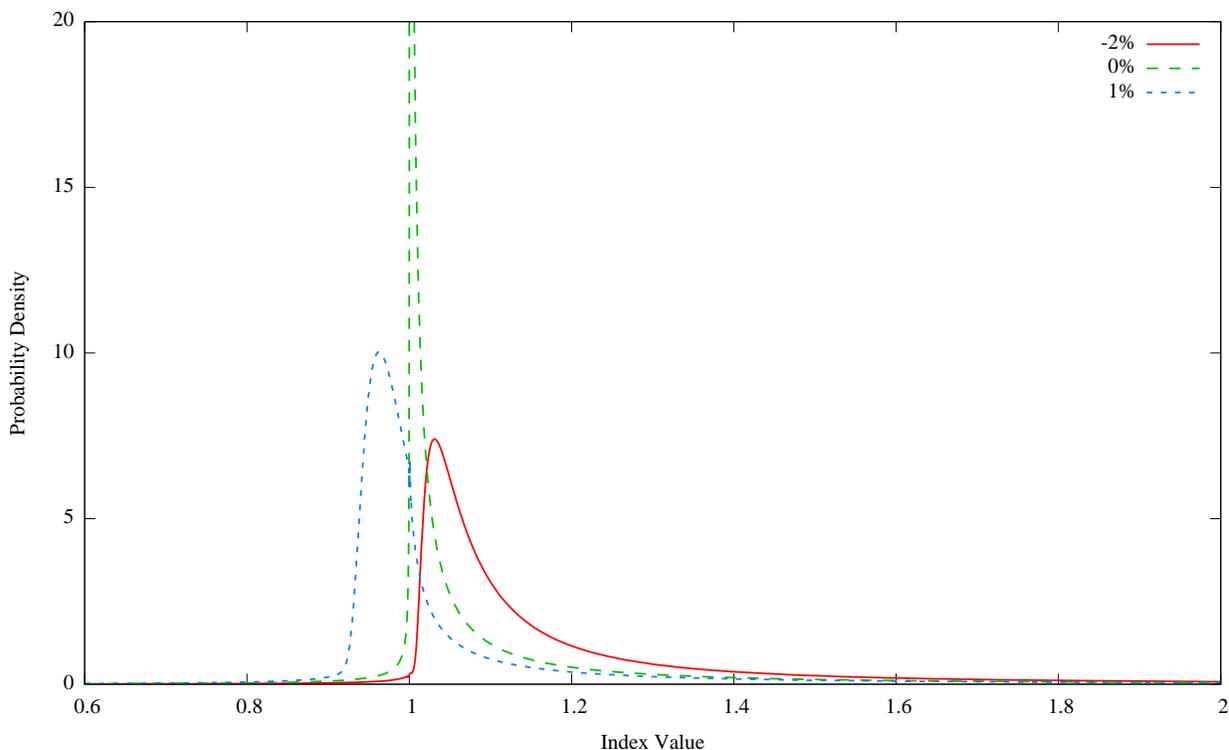}
\caption{\emph{Probability Density of the CPPI strategy depending on the threshold spread.}}
\label{fig-threshold}
\end{figure}

For positive spreads, the effect is the opposite: the threshold starts below its natural value and goes up with time to reach the same level at the CPPI maturity. As the CPPI distribution tends to concentrate just above the threshold value, this results in the threshold progressively \emph{encroaching} some part of the distribution and produces large amounts of small losses as shown in figure \ref{fig-threshold}.
This effect could be compensated by introducing a cushion limit. Thus the effect of increasing the exposure at the beginning of the CPPI life would remain without this mechanical increase of the gap proportion. However increasing the exposure automatically increases the risk of large losses: the conditional loss increases also as shown in table \ref{tab-threshold}.

\begin{table}[htbp]
\centering
\begin{tabular}{|l|r|r|r|}
\hline
Spread & Gap Prop. & Cond. Loss & Expect. Loss \\
\hline
-2\,\%  & 1.66\,\%  & 35.61\,\% & 0.592\,\% \\
-1\,\%  & 2.47\,\%  & 31.52\,\% & 0.779\,\% \\
0\,\%   & 5.71\,\%  & 18.41\,\% & 1.052\,\% \\
0.5\,\% & 48.16\,\% &  3.87\,\% & 1.862\,\% \\
1\,\%   & 59.33\,\% &  5.41\,\% & 3.212\,\% \\
\hline
\end{tabular}
\caption{\emph{Gap proportion, conditional loss and expected loss depending on the threshold spread.}}
\label{tab-threshold}
\end{table}

When a linear threshold is chosen, depending on the initial level both regimes can exist successively: the linear threshold can grow faster than the natural threshold and slower at the end. The effect of both regimes will partially compensate each other.

\FloatBarrier

\pagebreak

\section{Conclusion}
We have introduced an efficient scheme for the pricing of CPPI strategies and options. Instead of following the underlying spot process as in a Monte-Carlo pricing, only the CPPI strategy is considered. CPPIs are very sensitive to tails events; these are correctly handled numerically so that loss estimates can be computed for various underlying process. CPPI value distribution depends mainly on the mean volatility and on the left tail of the short term distribution for the underlying. With pure Brownian motion, huge volatility levels are necessary to get some gap risk: deeply Out-The-Money forward volatility controls this gap risk. On the other side, the average evolution of a CPPI depends on average volatility until maturity. These two parameters should be controlled separately.
Introducing jumps is the simplest way to handle this problem in a homogeneous model. We performed some analysis for exponentially distributed jumps, other jump distribution can be used as well. As the pricing depends on forward volatility, a direct and natural extension would be to introduce stochastic volatility through a few volatility regimes (with jumps depending on the volatility level). This would be useful for CPPI management as the potential gap size increases with the leverage.

The numerical analysis presented in this paper give some indications on the effect of various features aimed at reducing the gap risk. The most efficient features are the ones which reduce the leverage of the structure. Other techniques, such as setting a cushion limit (where the CPPI is completely de-leveraged before reaching the floor), only reduce the diffusive risk which is already almost null.

We have not addressed the important question of hedging, which is beyond the scope of this work. The hedging of the regular part of the CPPI, where the threshold is not broken, is quite simple: the delta of the strategy is more or less equal to the risky asset weighting dictated by the strategy definition. The difficult part is gap risk hedging which requires instruments sensitive to large downward jumps of the underlying process. If such instruments exist, the pricing scheme described here can be used to compute hedging ratios for optimal hedging and get estimators of the residual risk.

\section*{Acknowledgements}
We warmly thank Martial Millet for numerous stimulating discussions and for his constant availability. We also thank Erwan Curien for giving us the opportunity to work on this subject. Both of them are also acknowledged for a careful reading of this paper and useful comments. This work was supported by Sophis Technology.

\pagebreak

\begin{appendices}

\section{Diffusion models}
\label{section-model}

\subsection*{Black-Scholes}

If there are no dividends, under the $T$-forward risk neutral measure, the forward price of $S$ at maturity is a martingale
\begin{equation*}
\frac{\mathrm{d}F}{F} = \sigma \, \mathrm{d}W
\end{equation*}
This leads to the finite variations
\begin{equation*}
\frac{F_{i+1}}{F_i} = e^{- \frac{1}{2} \int \sigma^2 \, \mathrm{d}t + \int \sigma \, \mathrm{d}W}
\end{equation*}

With constant volatilities the cumulative and the (conditional) expected value of the variations of $F$ up to some strike $K$ are given by the usual formulas:
\begin{eqnarray*}
\mathbb{P}\!\left[ \frac{F_{i+1}}{F_i} < K \right] &=& \mathcal{N}\!\left( d_+ \right)
\\
\mathbb{E}\!\left[ \frac{F_{i+1}}{F_i} \mathbbm{1}_{\frac{F_{i+1}}{F_i} < K } \right] &=& \mathcal{N}\!\left( d_- \right)
\end{eqnarray*}
with
\begin{equation*}
	d_\pm = \frac{1}{\sigma \sqrt{t_{i+1} - t_i}} \left( \ln K \pm \frac{1}{2} \sigma^2 (t_{i+1} - t_i) \right)
\end{equation*}
where $\mathcal{N}$ denotes the cumulative function of a standard normal law.

\subsection*{Jump-diffusion}

Jumps can be added to the diffusion process in order to take into account large deviations in short timescale.
The jump-diffusion model introduced by Kou \cite{kou:jdm} has three independent components: a Black-Scholes diffusion, downward jumps and upward jumps both exponentially distributed with distinct parameters. The asset price is the exponential of the sum of the three components. As a consequence, the distribution of the asset price logarithm is the convolution of three distributions. It can be efficiently computed by Fourier transforms.

The diffusion is given by the following equation (again without dividends and under the $T$-forward risk neutral measure)
\begin{equation*}
\mathrm{d} \ln F = \gamma \, \mathrm{d}t + \sigma \, \mathrm{d}W + \mathrm{d}J_+ + \mathrm{d}J_-
\end{equation*}
$\mathrm dW$ is a standard Brownian motion, $\mathrm d J_\pm$ are jump processes with intensity $\lambda_\pm$ and mean jump size $\eta_\pm$. The jumps' size distributions are
\begin{equation*}
\nu_\pm(z) = \frac{1}{\eta_\pm}e^{\mp z/\eta_\pm} \mathbbm{1}_{\pm z >0}
\end{equation*}
The Brownian process $dW$, jump arrival times and jump sizes are all independent one from another. To ensure that $F$ is a martingale, the parameter $\gamma$ must be set to
\begin{equation*}
\gamma = - \frac{\sigma^2}{2} - \frac{\eta_+}{1-\eta_+} \lambda_+ + \frac{\eta_-}{1+\eta_-} \lambda_-
\end{equation*}

Moreover, the mean size of positive jumps $\eta_+$ must be lower than 1, i.e.\ the asset price cannot double or more during an infinitesimal period of time, in order to define a well behaved process.\footnote{Similarly, the variance of the process $S$ is infinite for $\eta_+ \geq \frac{1}{2}$ and, more generally, the $n^{\text{th}}$ moment will diverge if $\eta_+ \geq \frac{1}{n}$.}

Assuming constant coefficients for simplicity, the distribution $\phi$ of $\frac{F_{i+1}}{F_i}$ can be computed by Fourier transform from the characteristic exponent $\psi(u)$:
\begin{equation*}
\phi(x) = \frac{1}{2\pi} \int \mathrm{d}u \, e^{-iux} e^{\psi(u) (t_{i+1} - t_i)}
\end{equation*}
with
\begin{equation*}
\psi(u) = i \gamma u - \frac{\sigma^2 u^2}{2} + \frac{iu\eta_+}{1-iu\eta_+} \lambda_+ - \frac{iu\eta_-}{1+iu\eta_-}  \lambda_-
\end{equation*}

Numerically, we make use of the cumulative function and the expected value of the variation of $F$ up to some strike $K$. They can be computed directly by FFT techniques, without integrating the distribution. However one has to introduce an auxiliary distribution with known Fourier transform, such as a Gaussian distribution and compute corrections to this distribution.

A cumulative function has limit $0$ in $-\infty$ and $1$ in $+\infty$: its Fourier transform is not well-defined as a function. Subtracting the cumulative distribution of a gaussian distribution, the function which has to be transformed has limit zero in both  $-\infty$ and  $+\infty$ and can be considered to live on a circle. FFT gives the Fourier transform of this difference of cumulatives which has to be added to the cumulative function of the gaussian distribution to recover the cumulative distribution of the jump process. In fact, the difference function obtained by FFT is obtained on a grid; a cubic spline interpolation is used to get values everywhere. We take a gaussian with same mean and variance as the process considered in order to reduce numerical errors.

The second function we make use of in the computation is the expected value of the variation of $F$ up to some strike $K$, $\mathbb{E}\!\left[ \frac{F_{i+1}}{F_i} \mathbbm{1}_{\frac{F_{i+1}}{F_i} < K } \right]$. It may be rephrased as a cumulative function on a Kou process with a different set of parameters and it can thus be obtained by the method just described. The new set of parameters is
\begin{equation*}
	\left( \begin{array}{c}
	\gamma \\
	\sigma \\
	\lambda_\pm \\
	\eta_\pm \end{array} \right) \longrightarrow
	\left( \begin{array}{c}
	\gamma' \\
	\sigma' \\
	\lambda'_\pm \\
	\eta'_\pm \end{array} \right) =
	\left( \begin{array}{c}
	\gamma + \sigma^2 \\
	\sigma \\
	\lambda_\pm (1 \mp \eta_\pm)^{-1} \\
	\eta_\pm (1 \mp \eta_\pm)^{-1} \end{array} \right)
\end{equation*}
In the case where both $\lambda_\pm$ vanish, $\eta_\pm$ are no longer relevant, this reduces to the usual Black-Scholes computation.

\section{Closed formulas}
\label{appendix-closed-formula}

In the following, we consider a time-homogeneous vanilla CPPI with equally spaced rebalancing dates: $t_0 < t_1 < \dots < t_n$ with $t_i - t_{i-1} = \tau$ for $i = 1, \dots, n$. From equation \eqref{eq-SF-oneperiod} and from the definition of the standard exposure \eqref{eq-standard-exposure}, we obtain
\begin{equation*}
X_{k+1} = m \left( X_k - 1 \right)^+ \frac{F_{k+1}}{F_k} - m \left( X_k - 1 \right)^+ + X_k
\end{equation*}
or, equivalently
\begin{equation*}
X_{k+1} - 1 = \begin{cases} \left( X_k - 1 \right) \left[ m \frac{F_{k+1}}{F_k} - m  + 1 \right], \qquad & \text{if } X_k > 1 \\
                            X_k - 1, \qquad                                                             & \text{if } X_k \leq 1 \end{cases}
\end{equation*}
This formulation is useful when seeking information on the sign of $X_k - 1$ which enables us to tell wether the CPPI fell below the threshold or not. Indeed if we denote by $S$ the (stochastic) index defined as
\begin{equation*}
S = \min \left\{ k \in \{ 1 \dots n \} \text{ such that } X_k \leq 1 \right\}
\end{equation*}
the previous relation could be rewritten as
\begin{equation}
X_{k+1} - 1 = \left( X_0 - 1 \right) \prod_{i=1}^{S \wedge k+1} \left( m \frac{F_{i}}{F_{i-1}} - m  + 1 \right)
\label{eq-simple-dyn}
\end{equation}

We define the local short fall probability as $\mathrm P^\text{LSF} = \mathbb{P} \left[ X_1 < 1 \, \big\vert \, X_0 > 1 \right]$ (recall that the problem is invariant by translation in time), we have
\begin{equation*}
\mathrm P^\text{LSF} = \mathbb{P} \left[ \frac{F_{1}}{F_0} \leq \frac{m-1}{m} \right]
\end{equation*}
The gap proportion (or short fall probability) is given in term of the local short fall probability as
\begin{equation*}
\mathrm P^\text{SF} = \mathbb{P} \left[ X_n < 1 \, \big\vert \, X_0 > 1 \right] = 1 - \left( 1 - \mathrm P^\text{LSF} \right)^n
\end{equation*}
The (undiscounted) price of a European put option with strike $G$ is given by
\begin{eqnarray*}
\text{Put}(t_0) & = & G \, \mathbb{E} \left[ \left( 1 - X_n \right) \mathbbm{1}_{X_n \leq 1} \right] \\
& = & G \Big( \mathbb{E} \left[ \left( 1 - X_n \right) \right] - \mathbb{E} \left[ \left( 1 - X_n \right) \mathbbm{1}_{X_n \leq 1} \right] \Big) = G \Big( 1 - X_0 - \mathbb{E} \left[ \left( 1 - X_n \right) \mathbbm{1}_{S = \infty} \right] \Big)
\end{eqnarray*}
From equation \eqref{eq-simple-dyn}, we have
\begin{equation*}
\mathbb{E} \left[ \left( 1 - X_n \right) \mathbbm{1}_{S = \infty} \right] = \left( 1 - X_0 \right) \mathbb{E} \left[ \mathbbm{1}_{S = \infty} \prod_{i=1}^{n} \left( m \frac{F_{i}}{F_{i-1}} - m  + 1 \right) \right]
\end{equation*}
where we can explicitly decompose the event $\left\{ S = \infty \right\}$ as
\begin{equation*}
\left\{ S = \infty \right\} = \bigcap_{i=1}^{n} \left\{ \frac{F_{i}}{F_{i-1}} > \frac{m-1}{m} \right\}
\end{equation*}
The increments being stationary and independent, we get
\begin{equation*}
\mathbb{E} \left[ \left( 1 - X_n \right) \mathbbm{1}_{S = \infty} \right] = \left( 1 - X_0 \right) A^n
\end{equation*}
with
\begin{eqnarray*}
A & = & \mathbb{E} \left[ \left( m \frac{F_{1}}{F_{0}} - m + 1 \right) \mathbbm{1}_{\frac{F_{1}}{F_{0}} > \frac{m-1}{m}} \right] = 1 - B \\
B & = & \mathbb{E} \left[ \left( m \frac{F_{1}}{F_{0}} - m + 1 \right) \mathbbm{1}_{\frac{F_{1}}{F_{0}} \leq \frac{m-1}{m}} \right] = m \, \mathbb{E} \left[ \frac{F_{1}}{F_{0}} \mathbbm{1}_{\frac{F_{1}}{F_{0}} \leq \frac{m-1}{m}} \right] + (1-m) \, \mathrm P^\text{LSF}
\end{eqnarray*}
The final result for the put reads
\begin{equation*}
\text{Put}(t_0) = G \left( 1 - X_0 \right) \left( 1-A^n \right)
\end{equation*}
When the CPPI is already started with $t_0 \leq t < t_1$, separating the contribution from the first period yields the following formula
\begin{equation*}
\text{Put}(t) = G \, \mathbb{E} \left[ \left( 1 - X_1  \right) \mathbbm{1}_{X_1 < 1} \right] + G \, \mathbb{E} \left[ \left( 1 - X_1  \right) \mathbbm{1}_{X_1 > 1} \right] \left( 1-A^{n-1} \right)
\end{equation*}
Using equation \eqref{eq-SF-oneperiod}, one gets
\begin{equation*}
	\mathbb P \left[ X_1 < 1 \, \big\vert \, X_0 > 1 \right] = \mathbb P \left[ \frac{F_1}{F_t} \leq K \right]
\end{equation*}
and
\begin{equation*}
	\mathbb{E} \left[ X_1 \mathbbm{1}_{X_1 < 1} \right] = w_0 X_0 \frac{F_t}{F_0} \mathbb{E} \left[ \frac{F_1}{F_t} \mathbbm{1}_{\frac{F_1}{F_t} \leq K } \right] + (1-w_0)X_0 	\mathbb P \left[ \frac{F_1}{F_t} \leq K \right]
\end{equation*}
where
\begin{equation*}
	K = \frac{F_0}{F_t} \frac{m - 1}{m}
\end{equation*}

\begin{figure}[htbp]
\centering
\includegraphics[width=\textwidth]{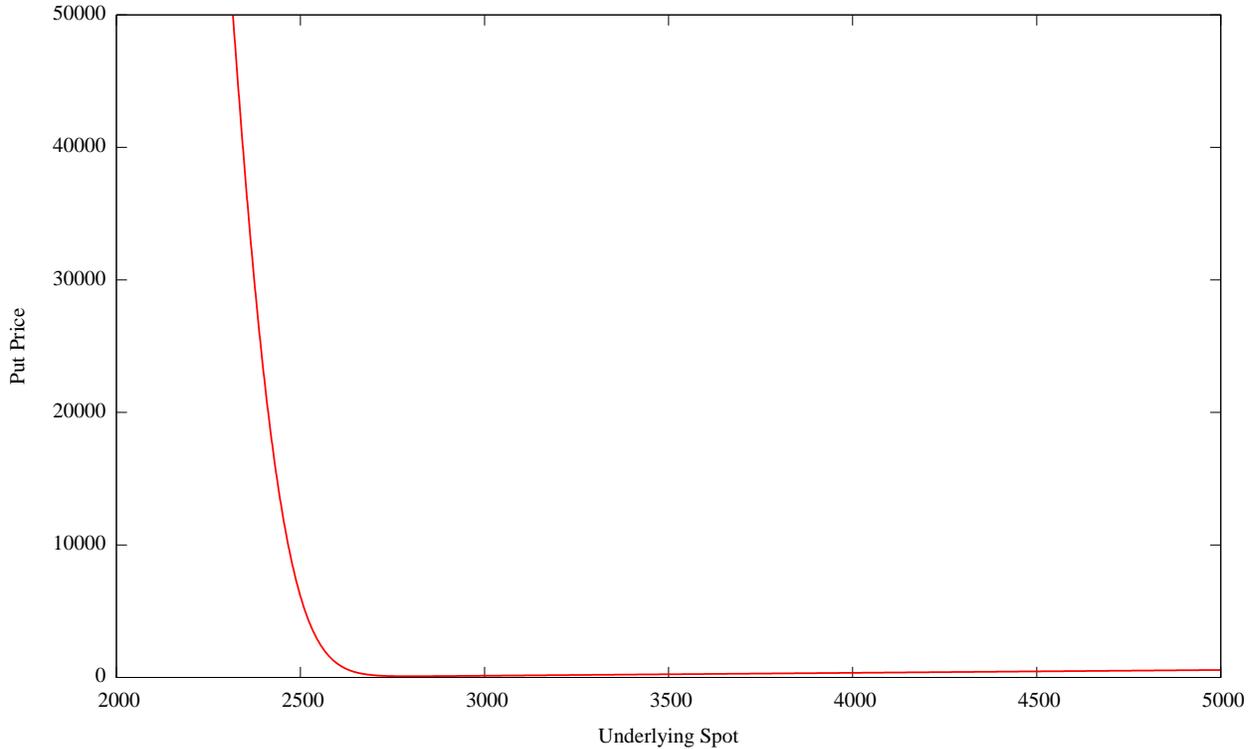}
\caption{\emph{Price of a CPPI Put depending on the underlying spot, for the vanilla CPPI described in section \ref{section-convergence}. The curve is asymptotically affine in both directions, with a much higher slope on the left, where the Put is in the money.}}
\label{fig-vanilla-price}
\end{figure}

In the case of a Black-Scholes model, one gets from appendix \ref{section-model} that $\mathbb P \left[ \frac{F_1}{F_t} \leq K  \right] = \mathcal N (d_+)$ and $\mathbb{E} \left[ \frac{F_1}{F_t} \mathbbm{1}_{\frac{F_1}{F_t} \leq K } \right] = \mathcal N (d_-)$. The undiscounted put price is then explicitly given by
\begin{equation*}
	\text{Put}(t) = G \Big( \mathcal N (d_+) - \mathcal N (d_-) \Big) A^{n-1} + G \Big( 1- w_0 X_0 \frac{F_t}{F_0} - (1-w_0)X_0 \Big) \left( 1 - A^{n-1} \right)
\end{equation*}
To illustrate the behaviour of the CPPI gap risk, we plot in figure \ref{fig-vanilla-price} the BS price of a vanilla CPPI put (the value of the guarantee) depending on the risky asset spot around the current spot value. Note the sign change in the Delta, with high Gamma.

\end{appendices}

\pagebreak
\bibliographystyle{alpha}
\bibliography{CPPI}

\begin{thebibliography}{BBM05}

\bibitem[Alb07]{albanese2006oma}
Claudio Albanese.
\newblock Operator methods, abelian processes and dynamic conditioning.
\newblock \url{http://www.level3finance.com/abelian.pdf}, 2007.

\bibitem[BBM05]{balder:ecs}
Sven Balder, Michael Brandl, and Antje Mahayni.
\newblock Effectiveness of {CPPI} strategies under discrete time trading.
\newblock \url{http://ssrn.com/abstract=898625}, 2005.

\bibitem[BK95]{boulier}
Jean-François Boulier and Anu Kanniganti.
\newblock Expected performance and risk of various portfolio insurance
  strategies.
\newblock In {\em Proceedings of the 5th AFIR International Colloquium}, pages
  1093--1124, Brussels, 1995.
\newblock
  \url{http://www.actuaries.org/AFIR/colloquia/Brussels/Boulier_Kanniganti.pdf%
}.

\bibitem[BP02]{bertrand2002pie}
Philippe Bertrand and Jean-Luc Prigent.
\newblock Portfolio insurance: the extreme value approach to the {CPPI} method.
\newblock {\em Finance}, 23(2):69--86, 2002.
\newblock \url{http://ssrn.com/abstract=299690}.

\bibitem[CT07]{cont:cpp}
Rama Cont and Peter Tankov.
\newblock Constant proportion portfolio insurance in presence of jumps in asset
  prices.
\newblock \url{http://ssrn.com/abstract=1021084}, 2007.

\bibitem[Kou02]{kou:jdm}
Steven~G. Kou.
\newblock A jump-diffusion model for option pricing.
\newblock {\em Management Science}, 48(8):1086--1101, 2002.
\newblock \url{http://ssrn.com/abstract=242367}.

\bibitem[Mer71]{merton1970oca}
Robert~C. Merton.
\newblock Optimum consumption and portfolio rules in a continuous-time model.
\newblock {\em Journal of Economic Theory}, 3(4):373--413, 1971.

\end{thebibliography}

\end{document}